\documentclass{ifacconf}

\usepackage{natbib}

\usepackage{graphicx} 
\usepackage{amsmath}

\usepackage{xcolor,soul,framed} 
\newtheorem{theorem}{Proposition}
\newtheorem{Theorem}{Lemma}
\colorlet{shadecolor}{yellow}
\usepackage{graphicx}
\usepackage{amsfonts}
\usepackage{multirow}
\usepackage{placeins}
\usepackage{nccmath}
\usepackage{cuted}

\usepackage{etoolbox}
\AfterEndEnvironment{strip}{\leavevmode}
\usepackage{float}
\usepackage{lipsum}
\usepackage{appendix}
\usepackage{array}
\usepackage{mdwmath}
\usepackage{mdwtab}
\usepackage{eqparbox}
\usepackage{multicol}
\usepackage{url}

\usepackage{natbib}  
\usepackage{lipsum,multicol}
\usepackage{dblfloatfix}

\begin{document}

\begin{frontmatter}

\title{Modeling  of Four-Winged Micro Ornithopters Inspired by Dragonflies} 

\author[First]{Oussama Sifour} 
\author[Second]{Soulaimane Berkane} 
\author[Third]{Abdelhamid Tayebi} 

\address[First]{Department of Computer Science
and Engineering, University of Quebec in Outaouais, Gatineau, QC, Canada. (e-mail: sifo01@uqo.ca).}
\address[Second]{Department of Computer Science
and Engineering, University of Quebec in Outaouais, Gatineau, QC, Canada. (e-mail: Soulaimane.Berkane@uqo.ca)}
\address[Third]{Department of Electrical Engineering,  Lakehead University, Thunder Bay, ON P7B 5E1, Canada. (e-mail: atayebi@lakeheadu.ca)}

\begin{abstract}                
In this paper, we present a full dynamical model of a four-winged micro ornithopter inspired by a dragonfly-type insect. The micro ornithopter is modeled as four articulated rigid body components (wings) connected to the main body via spherical joints. The dynamical model is derived using Lagrangian mechanics with intrinsic global coordinates, without relying on the common assumptions that neglect the wings-body interactions. Furthermore, the aerodynamic forces are modeled under the quasi-steady motion assumption without restricting the flapping frequency to be relatively high. This provides a full and elegant four-winged micro ornithopter model that captures the interaction between the body and the wings while avoiding the complexities and singularities associated with other coordinate representations (\textit{e.g.,} Euler angles). Simulation studies of the inertial effects of the relative motion between the different parts of the multibody system show the importance of considering the forces and torques, resulting from the wings-body interaction, in motion generation of these insects.
\end{abstract}


\end{frontmatter}

\section{Introduction}\label{sec1}
\vspace{-0.3cm}
Flapping wing unmanned aerial vehicles (FWUAVs), commonly known as ornithopters, have received an increasing attention over the last decade. The recent developments are encouraging, but the field is still in its infancy and considerable research efforts are needed to efficiently deploy these bio-inspired platforms. FWUAVs are particularly suitable for micro aerial vehicles (MAVs) applications where classical  aerial vehicles are inefficient, such as in search and rescue missions where these miniature autonomous vehicle can, for instance, fly through cracks in concrete to search for earthquake victims. 
In fact, FWUAVs enjoy some interesting features that cannot be found in fixed-wing and rotary-wing vehicles, such as energy efficiency, agility and miniaturization capacity. This has motivated to  study the aerodynamics and flight mechanics of insects and birds for clues that would help in the design of FWUAVs.
\vspace{-0.1cm}

The development of FWUAVs is often inspired from real birds and insects flights \citep{van2008insects}. One of the best flyers in the insect world is the dragonfly--a four-wing insect that has been extensively studied due to its astonishing flight capabilities. For example, using only its own body muscles, the dragonfly
can accelerate up to $3$g and reach a speed of $10 m/s$ \citep{3,5}, and is capable of generating an instantaneous lift five times greater than its weight \citep{6}. Each of its wings can be actuated independently \citep{7,8}, which provides additional degrees of freedom helping the insect to perform agile and complex flight maneuvers.
 The difficulty in modeling and controlling FWUAVS is mainly due to the aeroelastic phenomena and the intrinsic coupling between the wings and the main body of the vehicle. 
 Most of the existing mathematical models in the literature, for insect-inspired vehicles, rely on simplifying assumptions such as large flapping frequency, small body-wing mass ratio and neglecting the aerodynamic couplings between the different parts of the vehicle  \citep{9,10,11}.
 The resulting simplified models are often similar to single rigid-body systems with forces and torques depending on the flapping parameters. The coupling between the different parts of the multibody system (wings, tail, main body, etc), is particularly important for FWUAVs with large wings and relatively low flapping frequency. Some attempts to capture these coupling effects have been made in \citep{sridhar2020geometric,sun}. 
\vspace{-0.05cm}
It is known that insects generate aerodynamic forces through wings motion (flapping). These aerodynamic forces are often modeled using a quasi-steady assumption (the forces
generated by a  flapping wing are equal to the forces generated by the fixed wing at the same instantaneous
velocity and attitude of the wings blade). This approach was  mainly based on fixed-wing theory, but it has been used as a simple but robust modeling tool for flapping wings  for several decades \citep{1,14}. Later on, this approach was refined by the introduction of the lift and drag coefficients of the wings in flapping motion \citep{27,16}  to make it applicable for flapping wing vehicles. However, even at the fastest flight speeds, the quasi-steady aerodynamic interpretation seems inadequate to explain the extra lift produced by real insects flights. The importance of unsteady aerodynamic mechanisms for flapping insects flights has become widely recognized. Some numerical simulations of unsteady insect flight aerodynamics based on the finite element solution of the Navier–Stokes equations gave accurate results for the estimated aerodynamic forces \citep{17}. However, their implementation is unsuitable for control purposes since they require high processing power and, contrary to quasi-steady aerodynamics, they cannot be formulated as a function of the flapping parameters. In this work, using Langrange formalism, we propose a high-fidelity dynamical model for the dragonfly-like ornithopter, without relying on the high flapping frequency assumption and taking into consideration all the interaction forces and torques inherent to this multibody system. The dynamical model uses intrinsic global coordinates, mainly attitudes on the Special Orthogonal group of rotations,  which avoids the complexities and singularities associated with other coordinate representations (\textit{e.g.,} Euler angles). 
\vspace{-0.3cm}
\section {Notation}\label{sec2}
\vspace{-0.3cm}
We denote by $\mathbb{R}$ the set of reals and by $\mathbb{N}$ the set of natural numbers. We denote by $\mathbb{R}^{n}$ the $n-$dimensional Euclidean space. We use $\|x\|$ to denote the Euclidean norm of a vector $x \in \mathbb{R}^{n}$.  We denote by $\mathcal{F}=\{c,x,y,z\}$ an Euclidean 3-dimensional frame with center at $c\in\mathbb{R}^3$ and axes $\{x,y,z\}$ defining an orthonormal basis of $\mathbb{R}^3$. For a vector $x\in\mathbb{R}^n$, we denote by $\operatorname{sgn}(x):=\begin{bmatrix}
    \operatorname{sgn}(x_1),&\cdots,&\operatorname{sgn}(x_n)
\end{bmatrix}^\top$ the element-wise signum function of $x$. The special orthogonal group of order three is denoted by $\mathbb{S O}(3):=\left\{A \in \mathbb{R}^{3 \times 3}: \operatorname{det}(A)=\right.$
$\left.1, A A^{\top}=I\right\}$, where $I$ is the $3\times 3$ identity matrix. The unit vector $e_i$ denotes the $i$th column of the identity matrix $I$. The set $\mathfrak{s o}(3):=\left\{\Omega \in \mathbb{R}^{3 \times 3}: \Omega^{\top}=-\Omega\right\}$ denotes the Lie
algebra of $\mathbb{S O}(3).$ For each $z\in\mathbb{R}^n\setminus\{0\}$, we denote by $\mathbb{P}(z):=I-zz^\top\|z\|^{-2}$ the orthogonal projection operator. For $x, y \in \mathbb{R}^{3},$ the map $\hat{.}: \mathbb{R}^{3} \rightarrow \mathfrak{s o}(3)$
is defined as
$$
\hat{x}:=\left[\begin{array}{ccc}
0 & -x_{3} & x_{2} \\
x_{3} & 0 & -x_{1} \\
-x_{2} & x_{1} & 0
\end{array}\right].
$$
Then, we have $\hat{x}y:=x \times y$ where $\times$ is the vector cross-product on $\mathbb{R}^{3}$. 
Any rotation matrix $A\in\mathbb{SO}(3)$ can be parameterized by a unit vector $u\in\mathbb{R}^3$ and an angle $\theta \in \mathbb{R}$  through the exponential map  (Rodrigues formula)
$$
A=\exp \left(\theta \hat{u}\right):=I+\sin (\theta)\hat{u}+(1-\cos (\theta))\hat{u}^2.
$$
\vspace{-0.3cm}
\section{Mechanical Configuration}\label{sec3}
\begin{figure}[H]
    \centering
    \includegraphics[width=0.35\textwidth]{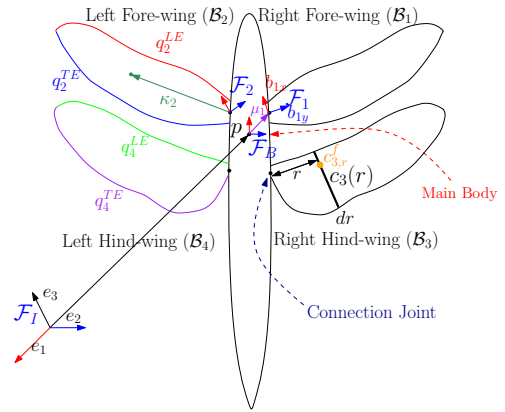}
    \caption{A schematic diagram of a four-winged ornithopter.}
    \label{fig1}
\end{figure}

\vspace{-0.3cm}
Let us consider a flapping wing micro aerial vehicle that  can translate and rotate in three dimensions as  multiple rigid bodies ($2$ fore-wings and $2$ hind-wings) connected to the main body via spherical joints that constrain the five rigid bodies to
remain in contact, see Fig. \ref{fig1}. 
We define six Euclidean frames: an inertial frame of reference $\mathcal{F}_I=\{0,b_{Ix},b_{Iy},b_{Iz}\}$, a body-attached
frame $\mathcal{F}_B=\{O_B,B_{bx},B_{by},B_{bz}\}$ for the main body, and a wing-attached frame $\mathcal{F}_i=\{O_i,b_{ix},b_{iy},b_{iz}\}$, $i\in \{1,\cdots,4\}$, where $O_B$ is the center of mass of the main body, and $O_i$ is the center of the joint connecting the the $i$th wing to the body. We use index $1$ for the right fore-wing, $2$ for the left fore-wing, $3$ for the right hind-wing, and $4$ for the left hind-wing. 

Let $A_B \in \mathbb{SO}(3)$ denote
the attitude matrix from the body-attached frame $\mathcal{F}_B$ to the inertial frame $\mathcal{F}_I$, and let $\Omega_B\in\mathbb{R}^3$ represent the angular velocity of the body-attached frame $\mathcal{F}_B$ with respect to the inertial frame $\mathcal{F}_I$ expressed in $\mathcal{F}_B$. Let $A_i \in \mathbb{SO}(3)$ denote the attitude matrix from the $i$th wing-attached frame $\mathcal{F}_i$ to the body-attached frame $\mathcal{F}_B$ and let $\Omega_i\in\mathbb{R}^3$ represent its angular velocity with respect to $\mathcal{F}_B$   expressed in $\mathcal{F}_i$. Now, let $\mu_i$, with $i\in \{1,\dots,4\}$, represent the vector from the origin of the body-attached frame $\mathcal{F}_B$ to the connection joint (center of frame $\mathcal{F}_i)$ expressed in $\mathcal{F}_B$. Let $\kappa_i$ represent the vector from the joint that connects the wing and the main body (center of frame $\mathcal{F}_i$) to the wing center of mass expressed in $\mathcal{F}_i$. Finally, let $p \in \mathbb{R}^3$ be the position vector of the origin of $\mathcal{F}_B$ expressed in $\mathcal{F}_I$.

Finally, the overall configuration of this multi-body system is denoted by $\mathfrak{g}:=(\mathfrak{g}_B,\mathfrak{g}_w)$, where $\mathfrak{g}_B:=(p,A_B) \in \mathbb{R}^3\times \mathbb{SO}(3)$  denotes the main body pose and $\mathfrak{g}_w:=(A_1,A_2,A_3,A_4) \in \mathbb{SO}(3)^4$ represents the four wings configurations (attitudes).  The group velocity is denoted by  $\xi:=(\xi_B,\xi_w)$, where $\xi_B:=(\dot{p},\Omega_B) \in \mathbb{R}^6$ is the main body group velocity and $\xi_w:=(\Omega_1,\Omega_2,\Omega_3,\Omega_4) \in \mathbb{R}^{12}$ represents the wings group velocity.

\vspace{-0.3cm}
\section {Quasi-Steady Aerodynamics}\label{sec4}
\vspace{-0.3cm}
Flapping-wings flights are more complicated than fixed-wings flights because of the structural movement and the resulting unsteady fluid dynamics. In conventional airplanes with fixed wings, the forward motion relative to the air causes the wings to produce lift. However, in a flapping-wings flight, the wings not only move forward relative to the air but also flap up and down and perform rotations around their roots. In this section, we will need to specify the geometry of the wings, as the resultant forces will depend on the shape of the wing. Then, we will use blade-element theory \citep{14} under the {\it quasi-steady aerodynamics assumption} to express the forces and torques generated through flapping motion. We rely on the quasi-steady assumption to simplify the forces and torques calculation. Intuitively, this assumption implies that these forces and torques are equivalent to those generated in steady motion at the same instantaneous velocity and the same angle of attack.
\vspace{-0.22cm}
\subsection{Wing Geometry}
\vspace{-0.2cm}
The dragonfly's wings are responsible  for its incredible flight performance and force generation. The shape of the fore-wing and hind-wing  of the dragonfly are often approximated by an elliptical function (tear-drop shape), see \citep{1}. However, as it is shown in \citep{24}, a more realistic representation of the wings' shape gives a better approximation of the aerodynamic forces generated 
by a real dragonfly, than the traditional tear-drop shape representation. A better approximation of the dragonfly's wings geometry can be obtained via a set of polynomials derived from the analysis of real photography images. We assume that each wing is a flat plate lying on the $b_{ix}-b_{iy}$ plane of $\mathcal{F}_i$ (no component on $b_{iz}$). Each wing's shape is described by two polynomial functions that represent the leading (upper bound) and the trailing (lower bound) edge, see Fig.~\ref{fig1}:
\begin{align}\label{eq1}
    q_{i}^{LE}(r):=\sum_{j=0}^{j=n} \lambda_{ij}^{LE}r^j,\quad
    q_{i}^{TE}(r):=\sum_{j=0}^{j=n} \lambda_{ij}^{TE}r^j,
\end{align}
where $r$ is the argument along the $b_{iy}$ axis and the $q_{i}^{LE}(r)$ and $q_{i}^{TE}(r)$ are the corresponding points locations of the trailing and leading edges, respectively, along the $b_{ix}$ axis. The polynomials coefficients $\lambda_{ij}^{LE}$ and $\lambda_{ij}^{TE}$  are calculated by fitting the points from the contour lines of the wings (obtained from real photography images) with polynomials of degree $n$ to minimize the mean squared errors while avoiding overfitting, see Section \ref{sec6} for a numerical example.

\vspace{-0.2cm}
\subsection{Aerodynamic Forces}
\begin{figure}[H]
\centering
\begin{tabular}{|c|c|}
\hline
\includegraphics[width=0.12\textwidth]{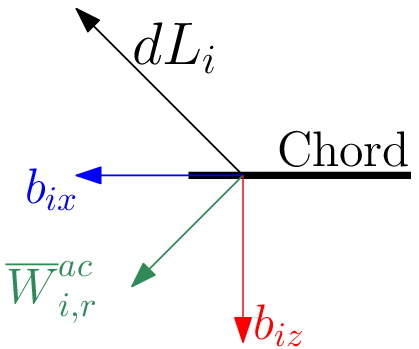}
     {case 1}
&
\includegraphics[width=0.12\textwidth]{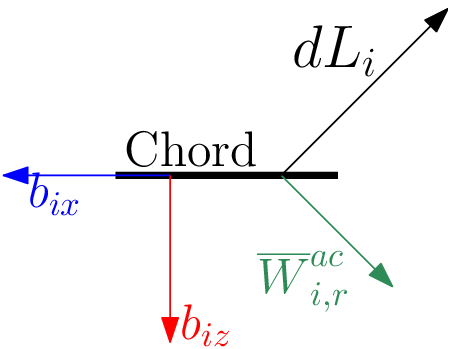}
     {case 2}
\\
\hline
\includegraphics[width=0.12\textwidth]{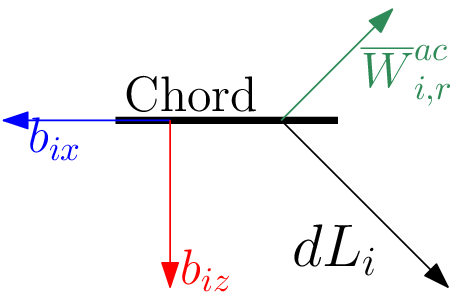}
     {case 3}
&
\includegraphics[width=0.12\textwidth]{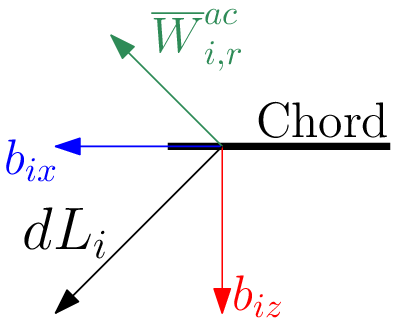}
     {case 4}
\\
\hline
\end{tabular}
\caption{Different cases of the direction of the lift with respect to the direction of $\overline{W}_{i,r}^{ac}$. Figure inspired from \citep{sridhar2020geometric}.}
\label{fig6}
\end{figure}

\vspace{-0.2cm}
To inspect the quasi-steady aerodynamic forces, let $dr$ be an infinitesimal wing segment, parallel to the $b_{ix}$ axis of the wing frame, and located at a distance $r$ from the wing root (Fig. 1), which is measured along the $b_{iy}$ axis  of the wing frame. Let its chord length be defined as $c_i(r):=q_i^{LE}(r)-q_i^{TE}(r)$ where $q_i^{LE}(\cdot)$ and $q_i^{TE}(\cdot)$ are defined as in \eqref{eq1}. 
%
\begin{Theorem}\label{lemma1}
    Let $\nu_{i}(r,\gamma):=(q_i^{LE}(r)-\gamma c_i(r))e_1+(-1)^{i+1}re_2$, with $\gamma \in[0,1]$, be the $\mathcal{F}_i$-coordinates of an arbitrary point on the chord. Then, the velocity of this point with respect to $\mathcal{F}_I$, expressed in $\mathcal{F}_i$, is given by
\begin{equation}
\begin{array}{rcl}
    W_{i,r,\gamma}(\mathfrak{g},\xi)&=&
(A_BA_i)^\top\dot{p}+A_i^\top\hat\Omega_B(\mu_{i}+A_i\nu_i(r,\gamma))\\
~&~&+ \hat \Omega_{i}\nu_i(r,\gamma).
\end{array}
\end{equation}
\end{Theorem}
\vspace{-0.15cm}
The first term of $W_{i,r,\gamma}(\mathfrak{g},\xi)$ is due to main body's translation motion, the second term is due the main body's rotation, and the last term is due to the wing's flapping. The proof of this lemma is given in appendix \ref{apen2}. 
Now, we proceed as in \citep{sridhar2020geometric} to determine the angle of attack. As per the blade-element theory \citep{14}, the aerodynamic force generated by the infinitesimal chord depends only on the $b_{ix}$, and $b_{iz}$ components. Therefore, we project the above velocity on the plane $b_{ix}-b_{iz}$ to obtain the {\it effective velocity}:
\begin{align}
    \overline{W}_{i,r,\gamma}(\mathfrak{g},\xi):=\mathbb{P}(e_2)W_{i,r,\gamma}(\mathfrak{g},\xi).
\end{align}
The state-dependent angle of attack of the chord, denoted $\alpha_{i,r}(\mathfrak{g},\xi)$, is the angle between the chord line (from the leading edge to the trailing edge),  and the above velocity $\overline{W}_{i,r,\gamma}$\footnote{We might drop the arguments of a given function whenever clear from context.}. Since the  angle of attack changes slightly chord-wise, we will define the angle of attack as the angle between the chord line and the velocity of the center of the chord $(\gamma= 1/2)$, and it is given by
\begin{align}
\alpha_{i,r}(\mathfrak{g},\xi):=\cos ^{-1}\left(\frac{e_{1}^\top \overline{W}_{i,r,\frac{1}{2}}(\mathfrak{g},\xi)}{\left\|\overline{W}_{i,r,\frac{1}{2}}(\mathfrak{g},\xi)\right\|}\right).
\end{align}%
\vspace{-0.17cm}
We consider the location of the state-dependent aerodynamic center, denoted $c^f_{i,r}(\mathfrak{g},\xi)$, as a function of the angle of attack  $\alpha_{i,r}$, and it is given by
\begin{align}\label{eqcif}
    c^f_{i,r}(\mathfrak{g},\xi):=\nu_{i}(r,\gamma_{ac}(\alpha_{i,r}(\mathfrak{g},\xi)).
\end{align}
where $\gamma_{ac}(\cdot):[0,\pi]\to[0,1]$ maps the angle of attack to the position of the aerodynamic center along the chord. The effective velocity of the aerodynamic center will be denoted by
$\overline{W}_{i,r}^{ac}(\mathfrak{g},\xi):=\overline{W}_{i,r,\gamma}(\mathfrak{g},\xi)$ with $\gamma=\gamma_{ac}(\alpha_{i,r}(\mathfrak{g},\xi))$. The magnitude of the lift and drag forces generated by the infinitesimal wing segment are given by
\vspace{-0.2cm}
\begin{align}
   \lVert dL_{i,r}(\mathfrak{g},\xi)\rVert&=\frac{1}{2}\rho\lVert \overline{W}_{i,r}^{ac}\rVert^2C_L(\alpha_{i,r})c_i(r)dr, \\
   \lVert dD_{i,r}(\mathfrak{g},\xi)\rVert&=\frac{1}{2}\rho\lVert \overline{W}_{i,r}^{ac}\rVert^2C_D(\alpha_{i,r})c_i(r)dr, 
\end{align}

where $\rho \in \mathbb{R}$ is the atmospheric density and $C_{L}, C_D:[0,\pi]\to\mathbb{R}$ are the lift and drag coefficients, which depend on the angle of attack $\alpha_{i,r}$. Following the same steps as \citep{sridhar2020geometric}, one can determine the infinitesimal lift and drag forces as follows. The direction of the lift is normal to both the velocity $\overline{W}_{i,r}^{ac}$ and the wing span-wise direction $b_{iy}$. As such, the
direction of the lift is along $ \pm e_2\times \overline{W}_{i,r}^{ac}$ in $\mathcal{F}_i$. Thus, we multiply the lift magnitude by the unit vector $(e_2\times \overline{W}_{i,r}^{ac})\|\overline{W}_{i,r}^{ac}\|^{-1}$. To solve the sign ambiguity induced by the flapping motion (up-stroke and down-stroke), we consider the four cases shown in Fig. \ref{fig6}.
\vspace{-0.1cm}
\vspace{-0.2cm}
From Fig. ~\ref{fig6}, one can notice that the direction of the lift force is in the direction of $e_2\times \overline{W}_{i,r}^{ac}$ if the first and third components of $\overline{W}_{i,r}^{ac}$ have the same sign (case 1 and case 3), otherwise it is in the direction of $ -e_2\times \overline{W}_{i,r}^{ac}$ (case 2 and case 4).
The direction of the drag force, however, is opposite to $\overline{W}_{i,r}^{ac}$. Finally, the corresponding aerodynamic forces and torques generated by the infinitesimal wing segment can be expressed in $\mathcal{F}_i$ as follows:
{\small\begin{align*}
    d L_{i,r}(\mathfrak{g},\xi) &=\frac{1}{2} \rho C_{L}(\alpha_{i,r}) c_i(r) \left\|\overline{W}_{i,r}^{ac}\right\| \operatorname{sign}(\bar{w}_{rx}^i \bar{w}_{rz}^i)(e_{2} \times \overline{W}_{i,r}^{ac})dr,\\
d D_{i,r}(\mathfrak{g},\xi) &=-\frac{1}{2} \rho C_{D}(\alpha_{i,r}) c_i(r)\left\|\overline{W}_{i,r}^{ac}\right\| \overline{W}_{i,r}^{ac} d r, 
\end{align*}}%
where $\bar{w}_{rx}^i$ and $\bar{w}_{rz}^i$ are the first and third components of $\overline{W}_{i,r}^{ac}$, respectively.
The infinitesimal lift and drag forces, which are applied at the aerodynamic center, also generate the following infinitesimal torque about the wing root
\begin{align}
\begin{split}
    d M_{i,r}(\mathfrak{g},\xi)=c_{i,r}^f(\mathfrak{g},\xi) \times\left(d L_{i,r}+d D_{i,r}\right) .
\end{split}
\end{align}
The total lift $L_{i}(\mathfrak{g},\xi)$, drag $D_{i}(\mathfrak{g},\xi)$, and torque $M_{i}(\mathfrak{g},\xi)$ generated on the $i$th wing, are obtained by integrating the above infinitesimal expressions span-wise for $r \in[0, l_i]$, where $l_i>0$ is the wing's length.
\begin{align}
    &L_{i}(\mathfrak{g},\xi):=\int_{0}^{l_i} d L_{i,r}(\mathfrak{g},\xi), \label{eq10}\\ 
    &D_{i}(\mathfrak{g},\xi):=\int_{0}^{l_i} d D_{i,r}(\mathfrak{g},\xi), \label{eq11}\\
    &M_{i}(\mathfrak{g},\xi):=\int_{0}^{l_i}d M_{i,r}(\mathfrak{g},\xi). \label{eq12}
\end{align}
Compared with the other models considering uniform forces over the wing \citep{forces}, this approach captures the span-wide variations of the aerodynamic forces, which are critical for FWUAVs with large wings flapping at a relatively low frequency.
\vspace{-0.25cm}
\section{Dragonfly Modeling using Lagrange-D'Alembert principle }\label{sec5}
\vspace{-0.25cm}
In this section, we will derive the expressions of the kinetic and potential energies. We will  use them along with the aerodynamic forces, detailed in Section \ref{sec4}, to derive a complete dynamical model for the four-winged ornithopter, using the Lagrange-D'Alembert principle \citep{lagrange}.

\vspace{-0.2cm}
\subsection{Kinetic and Potential Energies }
\vspace{-0.2cm}
The kinetic and the potential energies of the complete multibody system, denoted respectively by $T$ and $U$, are the sum of the kinetic and potential  energies of each rigid-body. They are given by
\begin{align}\label{eq:energies}
T=\sum_{j \in \{B,1,2,3,4\}} T_j, \quad U=\sum_{j \in \{B,1,2,3,4\}}U_j.
\end{align}
 The kinetic and the potential energies are  explicitly expressed in the following proposition.
\begin{theorem}\label{prop2}
The total kinetic energy can be expressed as follows:
\begin{align}
T(\mathfrak{g},\xi)&=\frac{1}{2}
\sum_{i=1}^4
\begin{bmatrix}
\dot{p} \\
\Omega_B \\
\Omega_{i}
\end{bmatrix}^\top \mathbf{J}_{i}(\mathfrak{g})\begin{bmatrix}
\dot{p} \\
\Omega_B \\
\Omega_{i}
\end{bmatrix}, \label{eq:T}
\\
U(\mathfrak{g})&=-\sum_{i \in \{B,1,2,3,4\}}m_{i} \text{g} e_{3}^\top\left(p+A_B\left(\mu_i+A_i\kappa_i\right)\right),\label{eq:U}
\end{align} 
with $\mu_B=\kappa_B=0$ and $\mathbf{J}_{i}(\cdot)$ is the symmetric matrix 
\begin{multline*}
\mathbf{J}_{i}(\mathfrak{g}):=
\begin{bmatrix}
\left(\frac{1}{4}m_B+m_{i}\right) I& 
\star & 
\star \\
m_{i}\left(\hat{\mu}_i+\widehat{A_i\kappa}_i\right) A_B^\top &\mathbf{J}_{i}^{[22]}
&\star\\
m_{i} \hat{\kappa}_{i} A_{i}^\top A_B^\top &
J_i A_i^\top+m_i \hat{\kappa}_i^\top A_i^\top \hat{\mu}_i & J_i
\end{bmatrix},
\end{multline*}%
 where \\
$\mathbf{J}_{i}^{[22]}:=\Big(A_i J_i A_i^\top-m_i \hat{\mu}_i^2+ m_i\hat{\mu}_i^\top \widehat{A_i \kappa_i}+m_i\widehat{A_i \kappa_i} \hat{\mu}_i^\top+\frac{1}{4}J_B\Big)$,  $m_B$ and $m_i$ are the mass of the main body and the connected bodies respectively, g represents the acceleration of gravity, $J_B \in \mathbb{R}^{3 \times 3}$   represents the constant inertia matrix of the main body about $\mathcal{F}_B$, and $J_i \in \mathbb{R}^{3 \times 3}$ represents the inertia matrix of the $i$th body about $\mathcal{F}_i$.

%
\end{theorem}
The proof of this proposition is given in appendix \ref{apen1}.
\vspace{-0.3cm}
\subsection{Lagrange-D'Alembert principle}
\vspace{-0.2cm}
This principle consists of a modification of Hamilton’s  principle to incorporate the effects
of external forces. These external forces may or may not be derivable from
a potential. This modification  states that the infinitesimal variation of the  integral action of $T-U$ over a
fixed time period equals  the work, denoted $\delta \mathcal{W}$, done by the external forces, corresponding to an infinitesimal variation of the configuration, during this same time period (also known as the virtual work of the external forces). Formally, for any $t_0\geq 0$ and $t_f\geq t_0$, we have
\begin{align}\label{alembert}
    \int_{t_{0}}^{t_{f}} \delta(T(t)-U(t)) d t=\int_{t_{0}}^{t_{f}} \delta \mathcal{W}(t) d t.
\end{align}
This version of the variational principle requires determining
the virtual work that corresponds to an infinitesimal variation of the configuration. Let $L:=T-U$ represent the Lagrangian and  let $F_i:=L_i+D_i$ represent the sum of forces acting on the $i$th wing, and 
let $\tau_{i} \in \mathbb{R}^{3}$ be the control torque exerted at the joint connecting the $i$th wing to the body, expressed in the body-attached frame. Following the developments in  \citep{25}, and using $a_{bj} \in \mathbb{R}^{3}$ and $a_{ij} \in \mathbb{R}^{3}$  to denote the $j$th column of $A_B\in \mathbb{SO}(3)$ and the $j$th column of $A_i\in \mathbb{SO}(3)$, respectively, for $i=1,2,3$, the Lagrange-d'Alembert principle leads to the equations stated in the following proposition:
\begin{theorem} \label{prop3}
  The Lagrange-d'Alembert principle \eqref{alembert} leads to the following equations:
{\small\begin{align}
&\frac{d}{d t}\left(\frac{\partial L}{\partial \dot{p}}\right)-\frac{\partial L}{\partial p}=\sum_{i=1}^{4} A_B A_{i} F_{i}, \label{eq17} \\
&\frac{d}{d t}\left(\frac{\partial L}{\partial \Omega_B}\right)+\hat{\Omega}_B \frac{\partial L}{\partial \Omega_B}+\sum_{j=1}^{3} \hat{a}_{bj} \frac{\partial L}{\partial a_{bj}}=\sum_{i=1}^{4} \hat{\mu}_{i} A_{i} F_{i}-\sum_{i=1}^{4} \tau_{i}, \\
&\frac{d}{d t}\left(\frac{\partial L}{\partial \Omega_{i}}\right)+\hat{\Omega}_{i} \frac{\partial L}{\partial \Omega_{i}}+\sum_{j=1}^{3} \hat{a}_{ij}  \frac{\partial L}{\partial a_{ij}}=M_{i}+A_{i}^{\top} \tau_{i}.\label{eq19}
\end{align}   }
\end{theorem}
The proof is given in appendix \ref{apen3}.
\vspace{-0.2cm}
\subsection{Full Dynamical Model}
\vspace{-0.2cm}
In this subsection, we use proposition 3 and the expression of $L=T-U$ to derive a dynamical model for the dragonfly. Using the kinetic and potential energies in the  previous Lagrange-D'Alembert equations, and defining $\tau=(\tau_1,\tau_2,\tau_3,\tau_4)\in \mathbb{R}^{12} $, one can derive the dynamical model as follows:
\begin{align}\label{eq28}
    &\mathbf{C}(\mathfrak{g}) \dot{\xi}+ \mathbf{D}(\mathfrak{g},\xi)\xi=F_{a}\left(\mathfrak{g},\xi\right)+H_c\left(\mathfrak{g}\right)\tau+F_{ \text{g} }(\mathfrak{g}).
\end{align}
with $\mathbf{D}(\mathfrak{g},\xi):=S\left(\xi\right)\mathbf{C}(\mathfrak{g})+\textbf{N}(\mathfrak{g},\xi)$, such that $S\left(\xi\right):=\mathrm{diag}\begin{bmatrix}0_{3\times 3} &\hat{\Omega}_B&\hat{\Omega}_1&\hat{\Omega}_2&\hat{\Omega}_3&\hat{\Omega}_4\end{bmatrix}$ and  $\mathbf{C}(\mathfrak{g})$  is a $18\times 18$ matrix given by
\vspace{-0.12cm}
{\small\begin{align}
 \mathbf{C}(\mathfrak{g}):=\begin{bmatrix}
mI& \displaystyle\sum_{i=1}^4\textbf{J}_i^{[12]}& \textbf{J}_{1}^{[13]} & \textbf{J}_{2}^{[13]} & \textbf{J}_{3}^{[13]} & \textbf{J}_{4}^{[13]}\\
\displaystyle\sum_{i=1}^4\textbf{J}_i^{[21]}  &\displaystyle\sum_{i=1}^4 \textbf{J}_i^{[22]}  & \textbf{J}_{1}^{[23]} & \textbf{J}_{2}^{[23]} & \textbf{J}_{3}^{[23]}& \textbf{J}_{4}^{[23]} \\
\textbf{J}_{1}^{[31]} & \textbf{J}_{1}^{[32]} & \textbf{J}_{1}^{[33]} & 0_{3 \times 3}&0_{3 \times 3}& 0_{3 \times 3}\\
\textbf{J}_{2}^{[31]} & \textbf{J}_{2}^{[32]} & 0_{3 \times 3} & \textbf{J}_{2}^{[33]}&0_{3 \times 3}&0_{3 \times 3}\\
\textbf{J}_{3}^{[31]} & \textbf{J}_{3}^{[32]} & 0_{3 \times 3} &0_{3 \times 3} & \textbf{J}_{3}^{[33]}&0_{3 \times 3}\\\textbf{J}_{4}^{[31]} & \textbf{J}_{4}^{[32]} & 0_{3 \times 3} &0_{3 \times 3} & 0_{3 \times 3} &\textbf{J}_{4}^{[33]}
\end{bmatrix},
\end{align}}%
where $m:=\sum_{i\in\{B,1,\cdots,4\}}m_i$, and $\textbf{J}_i^{[jk]}$ denotes the matrix block at the $j$th line and the $k$th row of the matrix $\textbf{J}_i$. The $3\times 3$ block components of the $18\times 18$ matrix $\textbf{N}(\mathfrak{g},\xi)$ are given below:
\begin{fleqn}
\begin{align*}
\small
\begin{split}
&\textbf{N}_{11}=0,\\
&\textbf{N}_{12}=\displaystyle\sum_{i=1}^4-m_iA_B\hat{\Omega}_B\left(\hat{\mu}_i+\widehat{A_i\kappa_i}\right)-m_iA_B\widehat{A_i\hat{\Omega}_i\kappa_i}\\
&\textbf{N}_{1(k+2)}=-m_kA_B\left(\hat{\Omega}_BA_k+A_k\hat{\Omega}_k\right)\hat{\kappa}_k,\\
\end{split}
\end{align*}
\vspace{-0.15cm}
\begin{align*}
\small
\begin{split}
&\textbf{N}_{21}=\displaystyle\sum_{i=1}^4-m_i\left(\hat{\mu}_i+\widehat{A_i\kappa_i}\right)\hat{\Omega}_BA_B^\top+m_i\left(\widehat{\hat{\mu}_i\Omega_B}+\widehat{\widehat{A_i\kappa_i}\Omega_B}\right) A_B^\top\\
\end{split}
\end{align*}
\vspace{-0.15cm}
\begin{align*}
\small
\begin{split}
&\textbf{N}_{22}= \displaystyle\sum_{i=1}^4A_i\hat{\Omega}_iJ_iA_i^\top-A_iJ_i\hat{\Omega}_i A_i^\top-m_i\left(\hat{\mu}_i\widehat{A_i\hat{\Omega}_i\kappa_i}+\widehat{A_i\hat{\Omega}_i\kappa_i} \hat{\mu}_i\right),
\end{split}
\end{align*}
\vspace{-0.15cm}
\begin{align*}
\small
\begin{split}
&\textbf{N}_{2(k+2)}=A_k\hat{\Omega}_kJ_k-m_k\hat{\mu}_kA_k\hat{\Omega}_k\hat{\kappa}_k,\\
&\textbf{N}_{31}=\displaystyle\sum_{i=1}^4-m_i\hat{\kappa}_i\left(A_i^\top\hat{\Omega}_B+\hat{\Omega}_iA_i^\top\right)A_B^\top+m_i\hat{\kappa}_iA_i\hat{\Omega}_BA_B^\top\\
&+m_i\widehat{\hat{\kappa}_i\Omega_i}A_i^\top A_B^\top,
\end{split}
\end{align*}
\vspace{-0.15cm}
\begin{align*}
\small
\begin{split}
&\textbf{N}_{32}=\displaystyle\sum_{i=1}^4-J_i\hat{\Omega}_iA_i^\top+m_i\hat{\kappa}_i\hat{\Omega}_iA_i^\top\hat{\mu}_i-\widehat{J_iA_i^\top\Omega_B}A_i^\top\\
&-m_i\hat{\kappa}_iA_i^\top\left(\hat{\Omega}_B\hat{\mu}_i-\widehat{\hat{\mu}_i\Omega_B}\right),\\
&\mathbf{N}_{j(k+2)}=\widehat{A_k^\top\Omega_B}^\top J_k+m_k\widehat{A_k^\top\hat{\mu}_k\Omega_B}\hat{\kappa}_k^\top,
\end{split}
\end{align*}
\end{fleqn}

\vspace{-0.15cm}
with $k\in \{1,\cdots,4\}$ and  $j\in\{3,\cdots,6\}$. The aerodynamic, gravitational, and torque control forces are given by
\begin{align*}
    F_a:=\begin{bmatrix}
    \displaystyle\sum_{i=1}^4A_BA_iF_i\\
    \displaystyle\sum_{i=1}^4\hat{\mu}_iA_iF_i\\M_1\\M_2\\M_3\\M_4
    \end{bmatrix},   H_c(\mathfrak{g}):=\begin{bmatrix}
        0&0&0&0\\-I&-I&-I&-I\\A_1^\top &0&0&0\\0&A_2^\top&0&0\\0&0&A_3^\top&0\\0&0&0&A_4^\top
    \end{bmatrix},
\end{align*}
\vspace{-0.35cm}
\begin{align}
    F_\text{g}:=\left[\begin{array}{c}
m \text{g} e_{3} \\
\displaystyle\sum_{i=1}^4m_{i} \text{g}\left(\hat{\mu}_i+\widehat{A_i\kappa_i}\right) A_B^\top e_{3} \\
m_{1} \text{g} \hat{\kappa}_{1}\left(A_{1}^\top A_B^\top e_{3}\right) \\
m_{2} \text{g} \hat{\kappa}_{2}\left(A_{2}^\top A_B^\top e_{3}\right) \\
m_{3} \text{g} \hat{\kappa}_{3}\left(A_{3}^\top A_B^\top e_{3}\right)\\
m_{4} \text{g} \hat{\kappa}_{4}\left(A_{4}^\top A_B^\top e_{3}\right)
\end{array}\right].
\end{align}

\vspace{-0.15cm}
This is a complete dynamical model, that provides the position, orientation, and wing dynamics of a four-winged ornithopter, with input torques applied at the joints connecting the wings to the body. 
\vspace{-0.25cm}
\begin{figure}[H]
    \centering
    \includegraphics[width=0.35\textwidth]{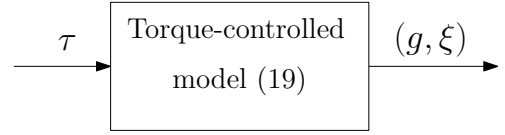}
    \caption{Torque-controlled model \eqref{eq28}.}
    \label{fig7}
\end{figure}
\vspace{-0.35cm}
\section{Reduced dynamical model via wings kinematics assignment}\label{sec6}
\vspace{-0.25cm}
Through a particular choice of the control inputs, one can generate wings motions mimicking real flapping-wing animals flights. This leads to a simplified model where the control inputs are the parameters related to the wings kinematics.

\vspace{-0.25cm}
\subsection{Desired Wing Flapping Kinematics}
\vspace{-0.25cm}
\cite{18} proposed a wing kinematic configuration for insects using Euler angles. This configuration minimizes energy in hovering flights, and captures several qualitative aspects of observed real insect flights. It can also be applied to other flight maneuvers (\textit{e.g.,} accelerating in different directions) as manipulating the three angles parameters generate aerodynamic forces in different directions. In the approach of \cite{18}, each wing's  attitude is given by three Euler angles: the flapping angle
$\phi_{i}(t)$, the deviation angle $\psi_{i}(t)$, and the pitching angle  $\theta_{i}(t)$, about the stroke frame. The stroke frame, denoted by $\mathcal{F}_{iS}=\{O_i,s_{ix},s_{iy},s_{iz}\}$, is obtained by rotating the wing frame $\mathcal{F}_i$  by a fixed angle $\beta_i \in \mathbb{R}$ about the body-frame $y$-axis. The corresponding wing attitude  for this kinematic configuration is taken from \citep{sridhar2020geometric},  and it is given by
\begin{align*}
    A_{i}=\exp \left(\beta_i \hat{e}_{2}\right) \exp ((-1)^{i+1}\phi_{i} \hat{e}_{1}) \exp ((-1)^i\psi_{i} \hat{e}_{3}) \exp \left(\theta_{i} \hat{e}_{2}\right),
\end{align*}
\vspace{-0.2cm}
with $i\in \{1,\dots,4\}$. The flapping angle is given by
\begin{align}\label{eq:phi}
   \phi_i(t)=\frac{\phi_{im}}{\sin ^{-1} \phi_{iK}} \sin ^{-1}\left(\phi_{iK} \cos (2 \pi f t+\phi_{ia})\right)+\phi_{i0},
\end{align}

where $f \in \mathbb{R}$ represents the flapping frequency in Hz,  $\phi_{im} \in \mathbb{R}$ is the amplitude, $\phi_{ia} \in \mathbb{R}$ is the phase, $\phi_{i0} \in \mathbb{R}$ is the offset, and $0<\phi_{iK} \leq 1$ determines the waveform shape (sinusoidal if $\phi_{iK} \rightarrow 0$, triangular if $\phi_{iK} \rightarrow 1$).

The pitch angle is given by the following function:
\begin{align}\label{eq:theta}
   \theta_i(t)=\frac{\theta_{im}}{\tanh \theta_{iC}} \tanh \left(\theta_{iC} \sin \left(2 \pi f t+\theta_{ia}\right)\right)+\theta_{i0},
\end{align}
where $\theta_{im} \in \mathbb{R}$ is the  amplitude, $\theta_{i0} \in \mathbb{R}$ is the offset, $\theta_{iC} \in(0, \infty)$ determines the waveform (sinusoidal when $\theta_{iC} \rightarrow 0$, step function when $\theta_{iC} \rightarrow \infty$), and $\theta_{ia} \in(-\pi, \pi)$
is the phase offset. The value of $\theta_{iC}$ is related to the duration of wing pitch reversal. Finally, the deviation angle is given by
\begin{align}\label{eq:psi}
  \psi_i(t)=\psi_{im} \cos \left(2 \pi \psi_{iN} f t+\psi_{ia}\right)+\psi_{i0},
\end{align}
where $\psi_{im} \in \mathbb{R}$ is the amplitude, $\psi_{i0} \in \mathbb{R}$ is the offset, and the parameter $\psi_{ia} \in(-\pi, \pi)$ is the phase offset. The parameter $\psi_{iN} \in\{1,2\}$, where $\psi_{iN}=1$ corresponds to one oscillation per flapping period, and $\psi_{iN}=2$ is for a figure-eight motion.

\vspace{-0.3cm}
\begin{figure}[H]
\begin{minipage}{0.3\linewidth}

  \includegraphics[width=1\textwidth]{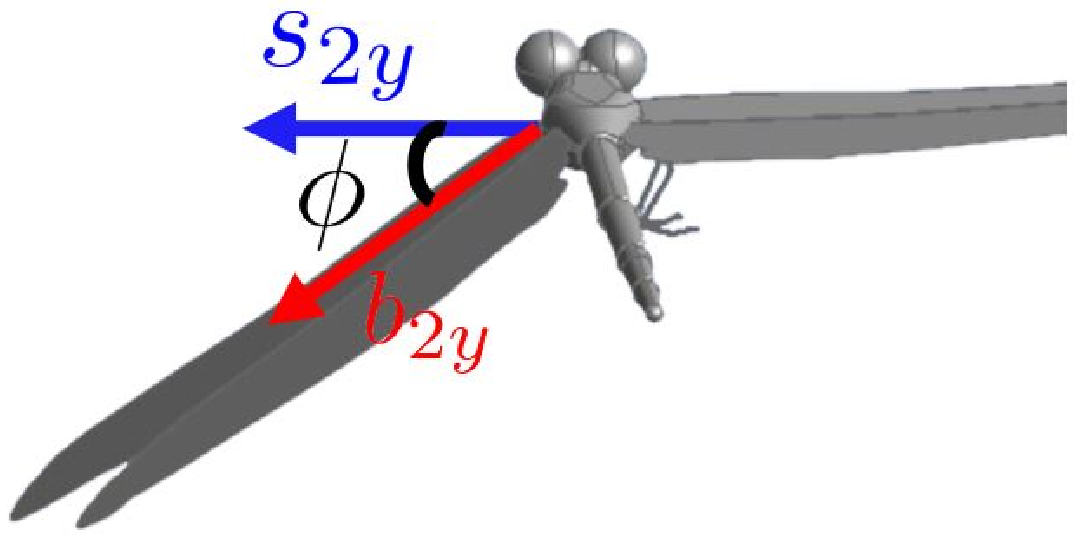}
  \label{fig:sfig14}
\end{minipage}%
\begin{minipage}{0.3\linewidth}

  \includegraphics[width=1\textwidth]{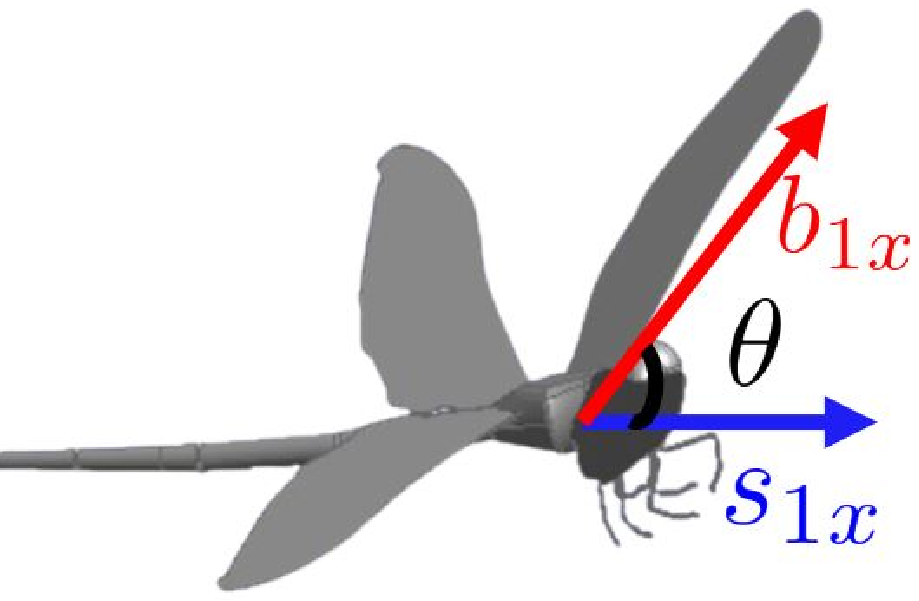}
  \label{fig:sfig26}
\end{minipage}
\begin{minipage}{0.3\linewidth}

  \includegraphics[width=1\textwidth]{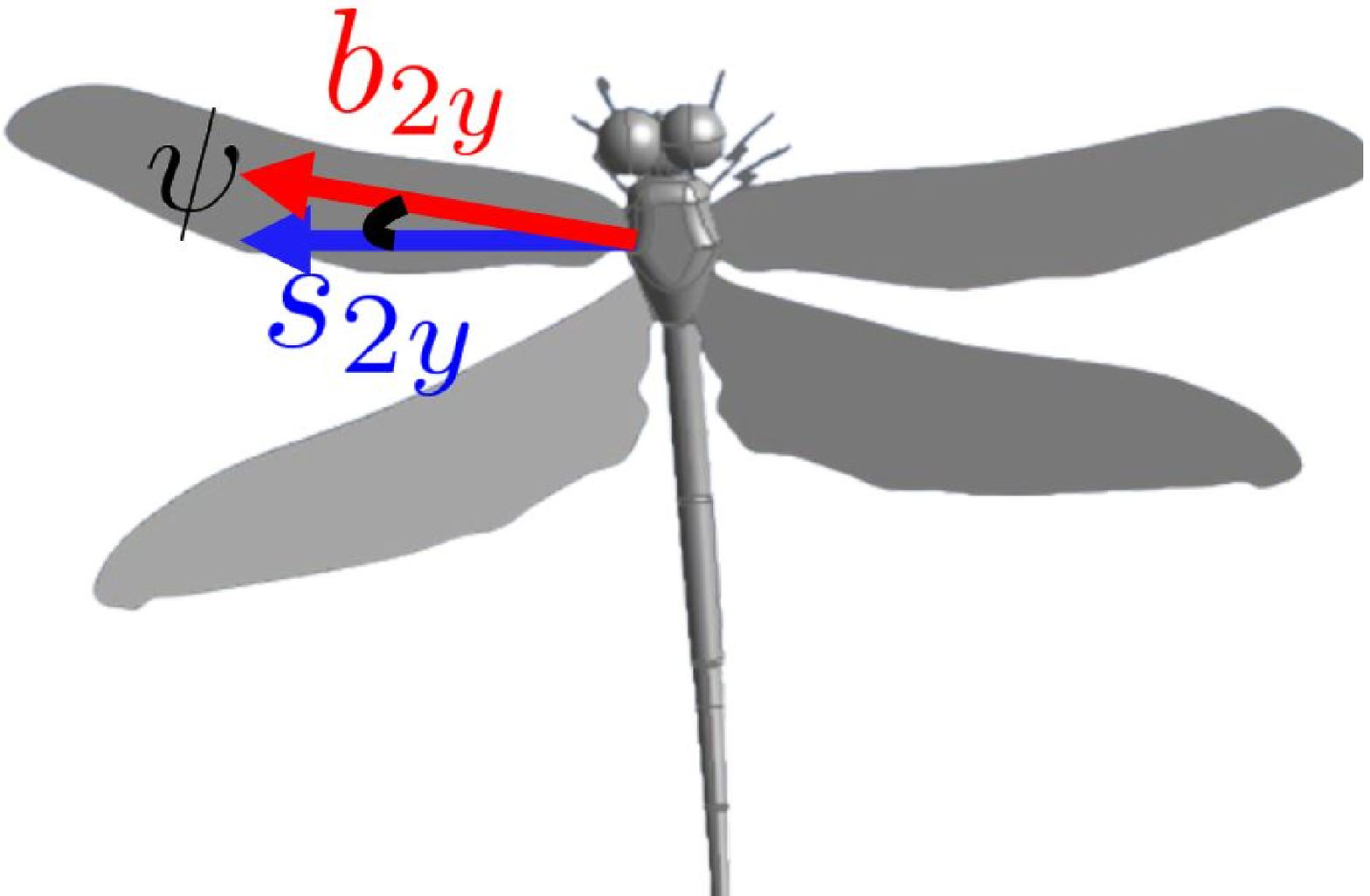}
  \label{fig:sfig27}
\end{minipage}
    \caption{Flapping, pitching, and deviation angles. Positive angles are measured from $\mathcal{F}_{iS}$  (in blue) to $\mathcal{F}_i$ (in red). }
\label{fig4}
\end{figure}
\vspace{-0.4cm}
For specific wings kinematics, we need to assign $13$ parameters per wing ($51$ parameters). These parameters are constrained for the dragonfly as in Table~\ref{constraint2}, see \citep{24}.

\vspace{-0.2cm}
\begin{table}[H]
$$
\begin{array}{|c|c|}
\hline \text { Parameter } & \text { range } \\
\hline f : \text {flapping frequency} & 30.0 : 45.00 \mathrm{~Hz}  \\
\phi_{im} : \text {flapping amplitude} & 30.0^{\circ} : 60.0^{\circ}  \\
\psi_{im} : \text {deviation amplitude}& 1.0^{\circ} : 20.0^{\circ}  \\
\theta_{im}: \text {pitching amplitude} & 1.0^{\circ} : 90.0^{\circ} \\
\phi_{i0} : \text {flapping offset}& -30.0^{\circ} : 30.0^{\circ} \\
\psi_{i0} : \text {deviation offset} & 5.0^{\circ}- : 30.0^{\circ}  \\
\theta_{i0} : \text {pitching offset} & -90.0^{\circ} : 90.0^{\circ}  \\
\psi_{ia} : \text {deviation phase} & -180.0^{\circ} : 180.0^{\circ}  \\
\phi_{ia} : \text {flapping phase} & -180.0^{\circ} : 180.0^{\circ}  \\
\theta_{ia} : \text {pitching phase} & -180.0^{\circ} : 180.0^{\circ}  \\
\phi_{iK} : \text {waveform shape} & 0.01 : 1.00  \\
\theta_{iC} : \text {waveform shape} & 0.01- : 5.00 \\
\beta_i : \text {stroke plane angle} & 5.0^{\circ} : 30.0^{\circ}  \\
\hline
\end{array}$$
    \caption{Parameters range for dragonfly wing kinematics.}
    \label{constraint2}
\end{table}

\vspace{-0.5cm}
The attitude kinematics of each wing is given by
\begin{align}\label{eq:Ai}
    &\dot{A}_i=A_i\hat{\Omega}_i,
\end{align}
with $i \in \{1,\cdots,4\}$. We consider the angular velocities of the wings  that are obtained from the time-derivatives of the Euler-angles equations \eqref{eq:phi}-\eqref{eq:psi} as follows \citep{sridhar2020geometric}:
\begin{align}\label{eq:Omega_i}
\Omega_{i}=\left[\begin{array}{ccc}
(-1)^{i+1}\cos \psi_{i} \cos \theta_{i} & 0 & (-1)^{i+1}\sin \theta_{i} \\
\sin \psi_{i} & 1 & 0 \\
(-1)^{i+1}\cos \psi_{i} \sin \theta_{i} & 0 & (-1)^{i}\cos \theta_{i}
\end{array}\right]\left[\begin{array}{c}
\dot{\phi}_{i} \\
\dot{\theta}_{i} \\
\dot{\psi}_{i}
\end{array}\right].
\end{align}

\vspace{-0.3cm}
\subsection{Reduced Dynamical Model }
\vspace{-0.3cm}
Let $(\mathfrak{g}^d_w,\xi^d_w)$ be a desired time-varying wings kinematics generated according to equations \eqref{eq:Ai}-\eqref{eq:Omega_i}, using a given parameters set $\Theta:=\left(f,\beta_i,\phi_{mi},\psi_{mi},\cdots\right)$ as described in Table II. Now, according to equation \eqref{eq19}, the control torque $ \tau_i$ is written as:
\begin{align}\nonumber
   \tau_{i}&=A_i\left(\frac{d}{d t}\left(\frac{\partial L}{\partial \Omega_{i}}\right)+\hat{\Omega}_{i} \frac{\partial L}{\partial \Omega_{i}}+\sum_{j=1}^{3} \hat{a}_{ij}  \frac{\partial L}{\partial a_{ij}}-M_i\right),\\
   &:=\mathbf{T}(\mathfrak{g}_B,\xi_B,\mathfrak{g}_w,\xi_w).
\end{align}
Now assuming that the inner loop is fast enough such that $(\mathfrak{g}_w,\xi_w)\approx(\mathfrak{g}^d_w,\xi^d_w)$, the control torques can be written as
\begin{align}
   \tau_{i}\approx \mathbf{T}(\mathfrak{g}_B,\xi_B,\mathfrak{g}_w^d,\xi_w^d)=:\mathbf{T}_\Theta^t(\mathfrak{g}_B,\xi_B),
\end{align}
where we have used the fact that $(\mathfrak{g}^d_w,\xi^d_w)$ is a function of parameters $\Theta$ and time $t$.
Replacing this torque expression in the second equation of Proposition \ref{prop3}, will allow us to write the translational, and rotational dynamics of the main body as follows: 
\begin{multline}\label{eq31}
\mathcal{C}^t_\Theta(\mathfrak{g}_B) \dot{\xi}_B+ \mathcal{D}^t_\Theta(\mathfrak{g}_B,\xi_B)\xi_B
   +\mathcal{V}^t_\Theta\left(\mathfrak{g}_B\right)=\mathcal{F}^t_\Theta\left(\mathfrak{g}_B,\xi_B\right).
\end{multline}
The matrices $\mathcal{C}_{\Theta}^t,\mathcal{D}_{\Theta}^t, \mathcal{V}_{\Theta}^t$ and $\mathcal{F}_{\Theta}^t$ and the development of this model are detailed in appendix \ref{apen4}. 
\vspace{-0.3cm}
\begin{figure}[H]
    \centering
    \includegraphics[width=0.45\textwidth]{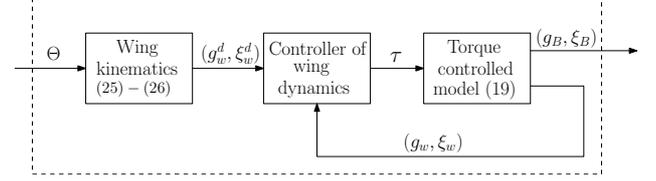}
    \caption{Parameters-controlled model.}
    \label{fig8}
\end{figure}
\vspace{-0.4cm}
  Grouping the forces and torques of the same nature together is another way to make this model ideal for designing control laws, and investigating wing-body interaction.
First, we write the aerodynamic forces and torques, which can be considered as control inputs in a tracking scenario. Second, we consider the inertial forces and torques of the wings due to body acceleration and rotation. Finally, we consider the inertial forces and torques of the wings due to wing motion (flapping). The model is given by  

\begin{align}\label{eq32}
    \begin{cases}
\dot{p}=v,\\
\dot{A}_B=A_B\hat{\Omega}_B,\\
m\dot{v}= m\text{g} e_3+F_c+F_B+F_w,\\
J_B\dot{\Omega}_B=\hat{\Omega}_BJ_B\Omega_B+\Gamma_c+\Gamma_B+\Gamma_w,
\end{cases}
\end{align}

where $v$ is the linear velocity of the body in the inertial frame, $F_c$ and $\Gamma_c$ are the aerodynamic forces and torques that can be used as a control inputs, $F_w$ and $\Gamma_w$ are the inertial forces and torques due to the flapping motion of the wings,  $F_B$ and $\Gamma_B$ are the inertial forces and torques due to the translational and rotational motion of the body. The expression of these forces are given in appendix \ref{apen5}. In most cases, the inertial forces due to the wing motion are neglected because of the wing mass $m_i$ being too small compared to the body mass $m_B$. In the case of the dragonfly, the wing mass $m_i$ represents around $3.5\%$ of the body mass $m_B$.  Moreover, neglecting both the inertial forces and torques due to the wing motion and body motion $F_B, F_w, \Gamma_B$ and $\Gamma_w$ results  in the widely used  model of a flying rigid body  \citep{gareth}, used in \citep{11,bee,19} to model insects flights. In the present work, we will take a closer look into the importance of these forces in near-hover scenarios.
\vspace{-0.3cm}
\section{Numerical Results}\label{sec7}
\vspace{-0.3cm}
In this section, we will investigate the effect of the inertial forces in a near-hover scenario. We will compare the trajectories performed by the dragonfly when neglecting the inertial forces versus when considering them. We will also compare the magnitudes of these inertial forces. 

\begin{figure}[ht]
    \centering
     {\small\renewcommand{\arraystretch}{1.6} 
   \hspace{-0.34cm} \begin{tabular}{c|c|c|c|c|c|}
    \cline{2-6}
       \multirow{7}{*}{\includegraphics[width=0.35\columnwidth]{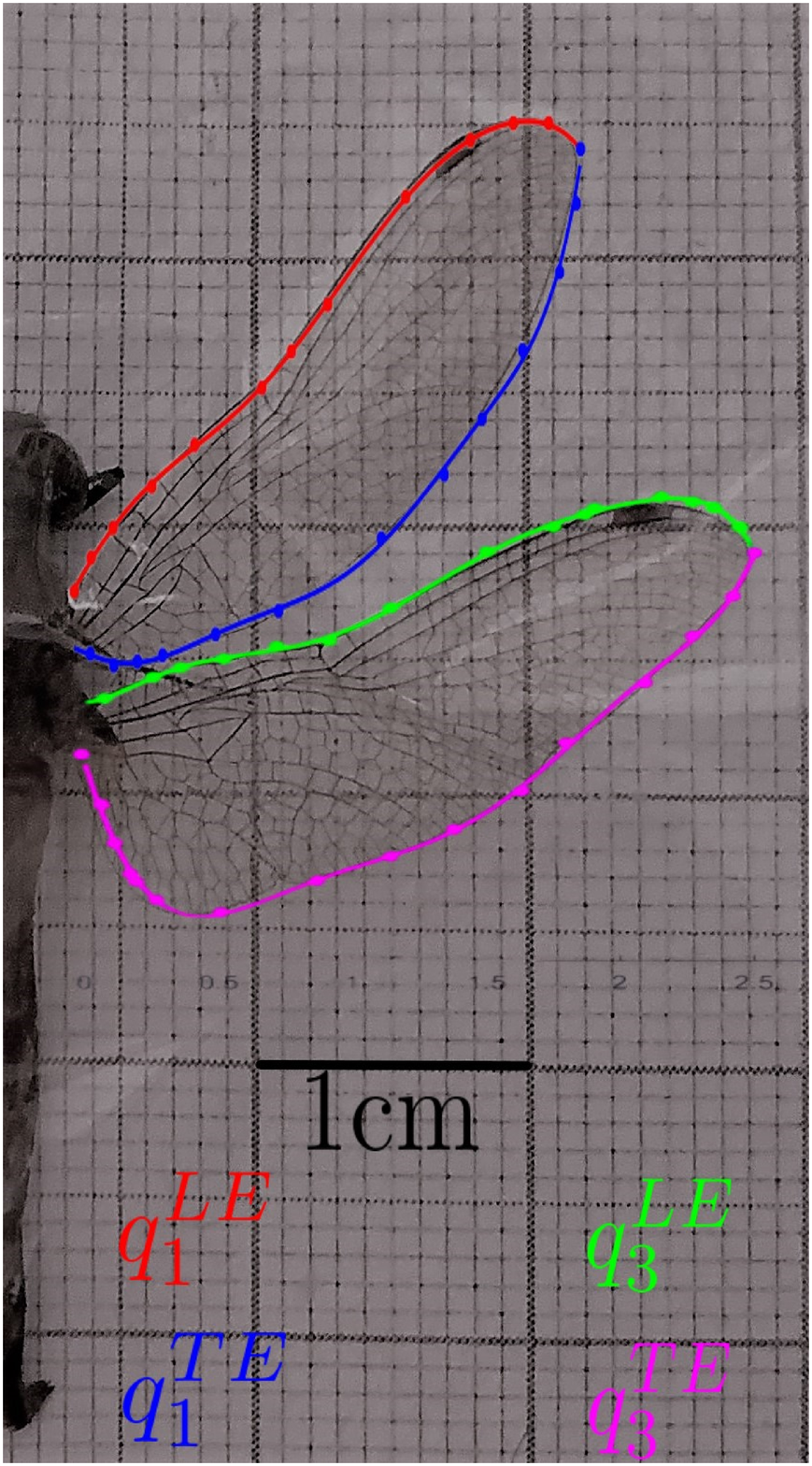}}\hspace{-0.18cm} 
&$j$&$\lambda_{1j}^{LE}$&$\lambda_{1j}^{TE}$& 
        $\lambda_{3j}^{LE}$& $\lambda_{3j}^{TE}$\\
        \cline{2-6}
          &0 & -0.873& 2.133& -0.214 & 0.183\\
         \cline{2-6}
           &1& 5.648 & -12.12 & 1.8389&   -1.372\\
           \cline{2-6}
           &2 &  -13.85 &  26.20 &  -6.153&   3.574\\
          \cline{2-6}
            &3 &  15.16 &  -26.25 & 9.981&   -3.016\\
            \cline{2-6}
              &4 &  -6.111 & -11.47 &  -7.857&  -2.210\\
             \cline{2-6}
              &5 & -0.579 & -0.728 & 2.625&   5.627\\
        \cline{2-6}
         &6 & 1.789 & -0.429 & 0.051&   -3.164 \\
         \cline{2-6}
          & 7 & 0.122 & -0.096  & 0.140&   -0.082\\
        \cline{2-6}
         &$R^2$ & 0.006 &  0.028 & 0.006&  0.012\\
        \cline{2-6}
    \end{tabular}}
   
    \vspace{0.5cm}
    \caption{Image of a dragonfly with coefficients
 of wings polynomials.}
 \label{fig22}
\end{figure}

First, it is necessary to describe the wings' shape, using the polynomial functions $q_{i}^{LE}$ and $q_{i}^{TE}$, since the aerodynamic forces depend on these polynomials, as shown in Section \ref{sec4}. Let us consider Fig. \ref{fig22} which shows  an image of a dead dragonfly insect. The polynomials' coefficients in equation \eqref{eq1} are calculated by fitting the points from the contour lines of the wings acquired by the algorithm in \citep{28} with $n=7$. The polynomials' coefficients are provided in the following table, and the polynomial functions are plotted in Fig. \ref{fig22}

Next, we take the location of the aerodynamic center $c_{i,r}^f$ similar to the location taken in \citep{dickinson}, \textit{i.e.,} the function $\gamma_{ac}(\alpha_{i,r})$ in equation \eqref{eqcif} is given by
\begin{align}
    \gamma_{ac}(\alpha):=0.82 \frac{|\alpha|}{\pi}+0.05.
\end{align}
The lift and drag coefficients $C_L(\alpha_{i,r})$, and $C_D(\alpha_{i,r})$ are functions of $\alpha_{i,r}$, and are taken similar to the coefficients proposed in \citep{27}:

\begin{align}
&C_L(\alpha_{i,r} ) = 0.225+1.58 \sin(2.13\overline{\alpha}_{i,r}^\circ - 7.20),\\
&C_D(\alpha_{i,r} ) = 1.92- 1.55 \cos(2.04\overline{\alpha}_{i,r}^\circ - 9.82).
\end{align}
with  $\overline{\alpha}_{i,r}^\circ=180\alpha_{i,r}/\pi$ if  $\alpha_{i,r} \leq \pi/2$ and $\overline{\alpha}_{i,r}^\circ=180(\pi-\alpha_{i,r})/\pi$ otherwise. 
To make the dragonfly perform a near-to-hover flight, we will need to find  suitable  parameters for the dragonfly's wings. This is challenging due to the complexity of the dynamics, and the relatively large number of the wing's kinematics parameters. This problem is addressed by performing an optimization to minimize a cost function, which is the error of the position and the velocity of the dragonfly from a reference hovering position. More specifically, this problem is formulated as follows. We define the cost
\vspace{-0.1cm}
$$\mathcal{J}=w_1\int_{t_{0}}^{T_{f}}\lVert p-p_{ref} \rVert^2 dt + w_2\int_{t_{0}}^{T_{f}} \lVert \dot{p}\rVert^2 d t,$$
    where $w_1,w_2 \in \mathbb{R}$, $T_f=10T$ is the flight time,   $T=1/f$ is the flapping period, and $p_{ref}=\begin{bmatrix}
        0& 0& 2
    \end{bmatrix}^\top$ is the hovering position reference. This is to minimize the error of the position and velocity to ensure that the dragonfly is hovering at $2m$ altitude. Next, as in \citep{tejaswi2021dynamics,sridhar2020geometric}, we assume that the dragonfly's body exhibits a pitching motion (oscillating around $e_2$), \textit{i.e.,} $A_B(t)=\exp(\Phi_B(t)\hat{e}_2)$, with $$\Phi_{B}(t)=\Phi_{B_{m}} \cos \left(2 \pi f t+\Phi_{B_{a}}\right)+\Phi_{B_{0}},$$
  where $\Phi_{B_{m}},\Phi_{B_{a}}$ and $\Phi_{B_{0}}$ are some parameters to be determined hereafter.   Replacing $A_B$ and $\hat \Omega_B=A_B^\top\dot A_B$ in \eqref{eq32}, the trajectory $p(t)$ can be obtained by integrating the translational dynamics (first and third equations of \eqref{eq32}). Finally, the optimization parameter is $\bar\Theta:=(\Theta,\Phi_{B_{m}},\Phi_{B_{a}},\Phi_{B_{0}})$.
 The constraints on the parameters are given in Table \ref{constraint2}, and the morphological parameters of the dragonfly like $J_B, J_i, \mu_i$  are taken to be similar to those of an actual insect, and are presented in appendix \ref{apen6}. This problem is carried out with the Genetic  Algorithm optimization \citep{21} implemented in MATLAB. The corresponding optimized parameters are summarized in Table \ref{tab2}, and the resulting trajectory is shown in  Fig. \ref{fig9}.
 
\begin{table}[h]
    \centering
    \small
    $$
\begin{array}{|c|c|c|}
\hline \text { Parameter } & \text { Fore-wing}_{1,2}  & \text { Hind-wing}_{3,4}  \\
\hline f & 35.6476 \mathrm{~Hz} & 35.6476 \mathrm{~Hz} \\
\phi_{im} & 58.42^{\circ} & 32.48^{\circ} \\
\psi_{im} & 11.16^{\circ} & 4.26^{\circ} \\
\theta_{im} & 1.43^{\circ} & 37.18^{\circ} \\
\phi_{i0} & 4.64^{\circ} & 28.49^{\circ} \\
\psi_{i0}& 26.49^{\circ} & 20.29^{\circ} \\
\theta_{i0} & -35.24^{\circ} & -1.83^{\circ} \\
\phi_{ia} & 0^{\circ} & 92.56^{\circ} \\
\psi_{ia} & -40.10^{\circ} & 29.37^{\circ} \\
\theta_{ia} & -98.82^{\circ} & -138.47^{\circ} \\
\phi_{iK} & 0.533 & 0.895 \\
\theta_{iC} & 2.394  &1.613 \\
\beta_i & 10.95^{\circ} & 23.53^{\circ} \\
\hline
\Phi_{B_{m}} & 0.37^\circ &\\
\Phi_{B_{a}}& -5.72^\circ& \\
\Phi_{B_{0}}& 0.434^\circ& \\
\hline
\end{array}$$
    \caption{Optimized parameters.}
    \label{tab2}
\end{table}

\begin{figure}[h]
    \centering
    \includegraphics[width=\columnwidth]{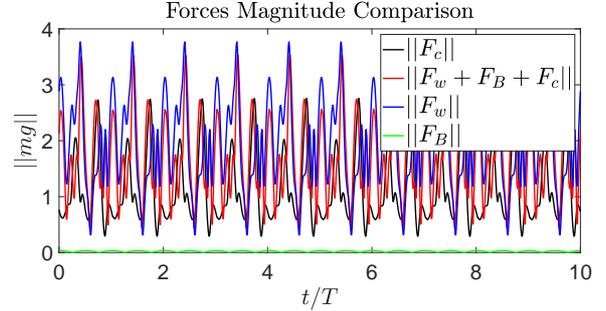}
    \caption{Forces magnitude comparison.}
    \label{fig10}
\end{figure}

\begin{figure}[h]
    \centering
    \includegraphics[width=\columnwidth]{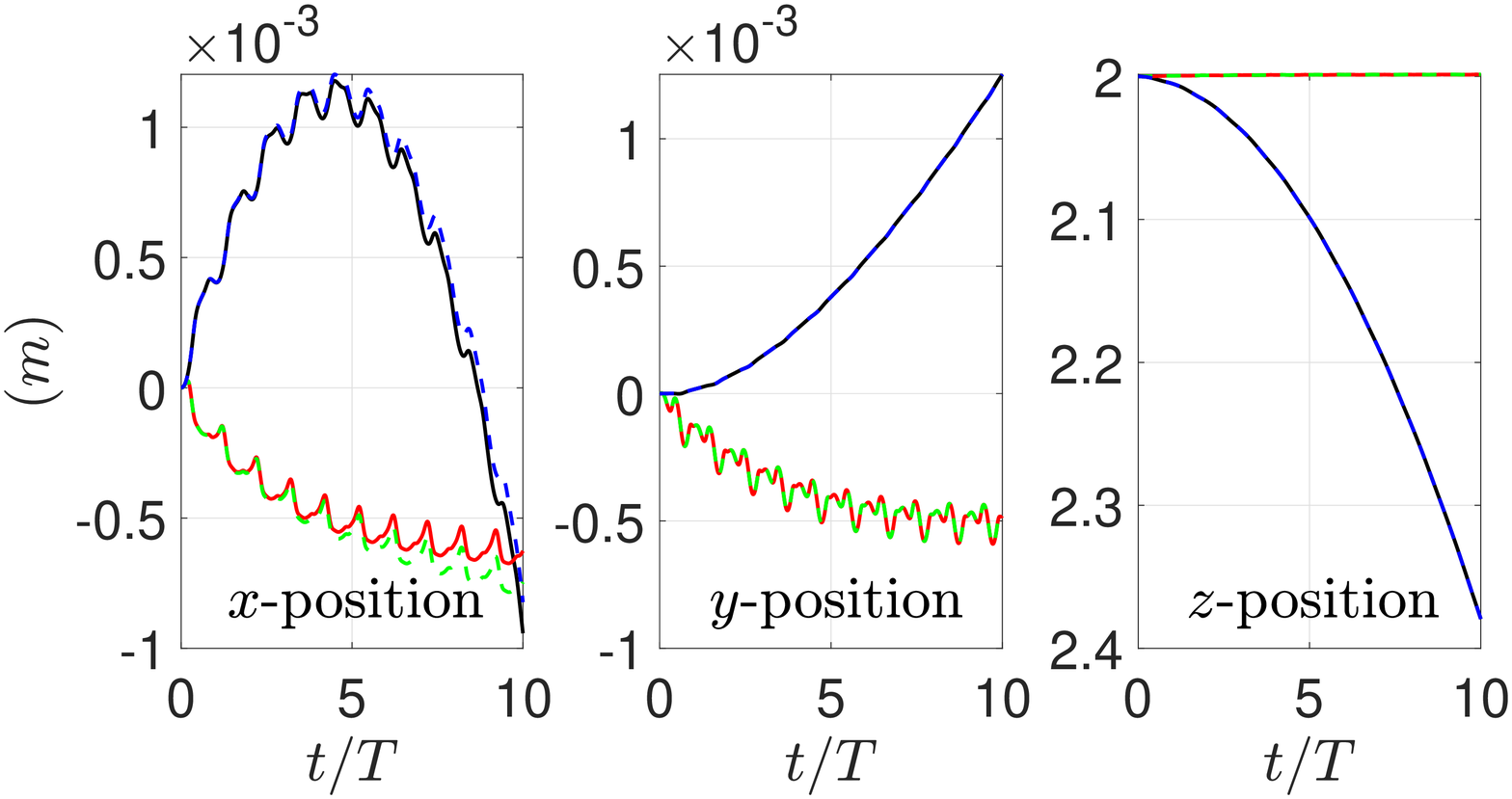}
    \caption{Position comparison for the full model in \eqref{eq32} (red), when neglecting $F_B$ and $F_w$ (black), when neglecting $F_B$ only (dashed green), and when neglecting $F_w$ only (dashed blue).}
    \label{fig9}
\end{figure}
\vspace{-0.18cm}
From Fig. \ref{fig9}, one can observe that neglecting the inertial forces due to body motion $F_B$,  results in a closer trajectory (dashed green line) to the full model trajectory (red line). This result is due to the small magnitude of the inertial forces due to body motion  $\left( \lVert F_B\rVert\right) $ as it is shown in Fig. \ref{fig10} in green lines. However, neglecting the inertial forces  due to the wing motion, results in a trajectory (dashed blue line in Fig. \ref{fig9}) relatively far from the full model trajectory (red line), and closer to the  flying rigid body model (\cite{11})  trajectory (black line). This result is due to the relatively important magnitude of the inertial forces due to the wing motion  $\left( \lVert F_w\rVert \right) $, as it is shown  in Fig. \ref{fig10} in blue line. Fig.\ref{fig9} and Fig.\ref{fig10} illustrate the importance of the wing-body interaction. The inertial forces caused by the wing motion are found to be quite important and should not be neglected. The inertial forces caused by the body motion are found to be negligible  in this scenario, due to the slow body motion compared with the wing motion. Considering the flying rigid body dynamical model as the one in \citep{11}, which is obtained by neglecting both inertial forces due to the body and wing motion, resulted in a different trajectory of the dragonfly. This trajectory is significantly far from its actual trajectory in view of the dragonfly's size.

\vspace{-0.3cm}

\section{Conclusion}\label{conc}
\vspace{-0.3cm}

A high-fidelity dynamical model of a four-winged micro ornithopter, inspired by a dragonfly-type insect, has been developed. The proposed model captures accurately the interaction between the body and the wings of the ornithopter, without relying on the common assumptions that neglect the wing masses and wing-body interactions. Additionally, we have used the quasi-steady aerodynamics assumption to model the span-wise aerodynamic forces generated by the infinitesimal wing chords, providing an elegant and accurate representation of the aerodynamics. Our simulation results demonstrate the importance of considering the inertial forces and the wing-body interactions in ornithopter dynamics. This highlights the contribution of our work in developing a full and elegant four-winged micro ornithopter model that captures the intricacies of the wing-body interaction. Finally, the configuration state of the proposed dynamical model evolves directly on the manifold $\mathbb{R}^3\times\mathbb{SO}(3)^5$ which avoids the complexities and singularities associated with other attitude representations.

\vspace{-0.25cm}
\begin{ack}
\vspace{-0.25cm}
This work was supported by the National Sciences and Engineering Research Council of Canada (NSERC), under the grants NSERC-DG RGPIN 2020-06270 and NSERC-DG RGPIN-2020-04759.
\end{ack}
\vspace{-0.3cm}
\bibliography{ifacconf}


\appendix

\section{Proof of Lemma \ref{lemma1}}\label{apen2}
Let us consider the root of one of the wings that is located at $p+A_B \mu_{i}$ with respect to the inertial frame. Thus, the $\mathcal{F}_I$-coordinates of an arbitrary point on the chord are given by
\begin{align}
\tilde{\nu}_{i,r,\gamma}(\mathfrak{g},\xi):=p+A_B \mu_{i}+A_B A_{i}\nu_i(r,\gamma).
\end{align}
Therefore, its linear velocity in $\mathcal{F}_{I}$ is given by
\begin{align}
    \frac{d}{dt}\tilde{\nu}_{i,r,\gamma}(\mathfrak{g},\xi)=&\dot{p}+A_B \hat{\Omega}_B \mu_{i}+A_B \hat{\Omega}_B A_{i} \nu_i(r,\gamma)\\ \nonumber
    &+ A_B A_i \hat{\Omega}_i\nu_i(r,\gamma),
\end{align}
which is expressed in $\mathcal{F}_{i}$ by left-multiplying $A_{i}^\top A_B^\top$ as:
\begin{align}
\begin{split}
    W_{i,r,\gamma}(\mathfrak{g},\xi):&=A_{i}^\top A_B^\top\frac{d}{dt}\tilde{\nu}_{i,r,\gamma}(\mathfrak{g},\xi)\\
    &=(A_BA_i)^\top \dot{p}+A_i^\top\hat\Omega_B(\mu_{i}+A_i\nu_i)+\hat\Omega_{i}\nu_i.
\end{split}
\end{align}    
\section{Proof of Proposition \ref{prop2}} \label{apen1}
The kinetic energy of the main body is the sum of the translational and rotational kinetic energies of the main body and is given by 
\begin{align}
    T_B=\frac{1}{2}m_B\lVert \dot{p}\rVert^2+\frac{1}{2}\Omega_B^\top J_B\Omega_B.
\end{align}
The kinetic energy of the each wing is the sum of the  translational and rotational kinetic energies and is given by
\begin{align}
    T_{i}=\frac{1}{2}m_i\lVert \dot{\kappa}_i^{I}\rVert^2+\frac{1}{2}\Omega_i^{I\top} \left(J_i+m_i\hat{\kappa}_i^2\right)\Omega_i^{I},
\end{align}
where $\kappa_i^{I}$ is the position of the center of mass of the $i$th wing in the inertial frame and is given by
\begin{align}
    \kappa_i^{I}=p+A_B\mu_i+A_BA_i\kappa_i,
\end{align}
and $\Omega_i^{I}$ is the angular velocity of the $i$th wing with respect to $\mathcal{F}_I$ expressed in $\mathcal{F}_i$ and is given by
\begin{align}
    \Omega_i^{I}=\Omega_i+A_i^\top\Omega_B.
\end{align}
The velocity of the center of mass of the $i$th wing is given by
\begin{align}
\dot{\kappa}_i^{I}=\dot{p}+A_B\hat{\Omega}_B\left(\mu_{i}+A_i\kappa_i\right)+A_BA_i\hat{\Omega}_i\kappa_i.
\end{align}
Substituting the expressions of $T_B$ and $T_i$ in \eqref{eq:energies}, we obtain the following expression of the total kinetic energy
\begin{align}
\begin{split}
    T=&\frac{1}{2}m_B\lVert\dot{p}\rVert^2+\frac{1}{2}{\Omega_B}^\top J_B\Omega_B+\frac{1}{2}\sum_{i\in \{1,\dots,4\}}m_i\Big( \lVert\dot{p}\rVert^2-\\
    &2\dot{p}^\top A_B\left(\hat{\mu}_i+\widehat{A_i\kappa_i}\right)\Omega_B-2\dot{p}^\top A_BA_i\hat{\kappa}_i\Omega_i- \\
&\Omega_B^\top\left(\hat{\mu}_i+\widehat{A_i\kappa_i}\right)^2\Omega_B+2\Omega_i^\top \hat{\kappa}_i^\top A_i^\top \left(\hat{\mu}_i+\widehat{A_i\kappa_i}\right) \Omega_B\Big)+\\
    &\frac{1}{2}\sum_{i\in \{1,\dots,4\}}\Big(\Omega_B^\top A_i\left(J_i+m_i\hat{\kappa}_i^2\right)A_i^\top \Omega_B+\\
    &2\Omega_i^\top \left(J_i+m_i\hat{\kappa}_i^2\right)A_i^\top\Omega_B+\Omega_i^\top J_i\Omega_i \Big),
    \end{split}
\end{align}
which can be rearranged as in \eqref{eq:T}. The potential energy of the main body is given by
\begin{align}
    U_{B}=-m_{B} \text{g} e_{3}^\top p,
\end{align}
and the potential energy of each connected body $\mathcal{B}_i, i \in \{1,\dots,4\} $ is given by 
\begin{align}
    U_{i}=-m_{i} \text{g} e_{3}^\top\left(p+A_B\left(\mu_i+A_i\kappa_i\right)\right).
\end{align}
The sum of potential energies gives \eqref{eq:U}.
\section{Proof of Proposition \ref{prop3}} \label{apen3}
 The left hand side of the equation $\delta \mathfrak{S}= \int_{t_{0}}^{t_{f}} \delta(T-U) d t$ can be written as follows:
    \begin{align*}
    \begin{split}
        &\delta \mathfrak{S}=\int_{t_{0}}^{t_{f}}\Bigg\{\frac{\partial L}{\partial \dot{p}} \delta \dot{p}+\frac{\partial L}{\partial p}  \delta p+\frac{\partial L}{\partial \Omega_B} \delta \Omega_B\\
        &+\frac{\partial L}{\partial A_B} \delta A_B+\sum_{i=1}^{4}\left(\frac{\partial L}{\partial \Omega_{i}} \delta\Omega_{i}+\frac{\partial L}{\partial A_{i}} \delta A_{i}\right)\Bigg\} d t
    \end{split}
    \end{align*}
     Replacing the infinitesimal variations of the orientation and the angular velocity by \citep{25} 
    \begin{align}\label{eqc1}
    \begin{split}
        &\delta A_B=A_B \hat{\eta}, \\
&\delta \Omega_B=\dot{\eta}+\hat{\Omega}_B \eta, \\
&\delta A_{i}=A_{i} \hat{\eta_i}, \\
&\delta \Omega_{i}=\dot{\eta}_{i}+\hat{\Omega}_{i} \eta_{i}.
\end{split}
    \end{align}
    where  $\eta:\left[t_{0}, t_{f}\right] \rightarrow \mathbb{R}^{3}$ is an infinitesimal variation  that vanishes at $t_{0}$ and $t_{f}$. The left-hand side of the equation \eqref{alembert} will result in the following equation:
    \begin{align}\label{lhs}
        \begin{split}
            &\delta \mathfrak{S}=\int_{t_{0}}^{t_{f}}\Bigg\{ \left(\frac{d}{d t}\left(\frac{\partial L}{\partial \dot{p}}\right)-\frac{\partial L}{\partial p}\right)\delta p+\\
            &\left(\frac{d}{d t}\left(\frac{\partial L}{\partial \Omega_B}\right)+\hat{\Omega}_B \frac{\partial L}{\partial \Omega_B}+\sum_{i=1}^{3} \hat{a}_{bi} \frac{\partial L}{\partial a_{bi}}\right)\eta+\\
            &\sum_{i=1}^{4}\left(\frac{d}{d t}\left(\frac{\partial L}{\partial \Omega_{i}}\right)+\hat{\Omega}_{i} \frac{\partial L}{\partial \Omega_{i}}+\sum_{j=1}^{3} \hat{a}_{ij}  \frac{\partial L}{\partial a_{ij}}\right)\eta_i \Bigg\} dt.
        \end{split}
    \end{align}
    The right hand side of \eqref{alembert} can be developed by considering an infinitesimal aerodynamic force $d F_i := dL_i+dD_i \in \mathbb{R}^3$ acting at the aerodynamic center location of $c_{i,r}^f \in \mathbb{R}^{3}$ of the wings. The location of the aerodynamic center is given by $\tilde{c}_{i,r}^f=p+A_B \mu_{i}+A_B A_{i}c_{i,r}^f$ in the inertial frame. Thus, the corresponding virtual work due to the aerodynamic force generated on one wing is given by
    \begin{align*}
\delta \mathcal{W}_i&=\int_{\mathcal{B}_{i}}\left(A_B A_{i} d F_i\right)^\top \delta\left(p+A_B \mu_{i}+A_B A_{i} c_{i,r}^f\right).
\end{align*}
Replacing the infinitesimal variations by their equations given in \eqref{eqc1}, the virtual work on one wing is given by
\begin{align*}
\begin{split}
\delta \mathcal{W}_i&=\left(A_B A_{i} \int_{\mathcal{B}_{i}} d F_i\right)^\top\left(\delta p+A_B \hat{\eta} \mu_{i}\right) \\
&+\left(\int_{\mathcal{B}_{i}}c_{i,r}^f \times d F_i\right)^\top \eta_{i}.
\end{split}
\end{align*}
We replace $dF_i$ and the integral over the wing body surface by its formula, to obtain
\begin{align}
\begin{split}
\delta \mathcal{W}_i&=\left(A_B A_{i} \int_{0}^{l_i} (d L_i+dD_i)\right)^\top\left(\delta p+A_B \hat{\eta} \mu_{i}\right) \\
&+\left(\int_{0}^{l_i} c_{i,r}^f \times (d L_i+dD_i) \right)^\top \eta_{i}.\label{eqwi}
\end{split}
\end{align}
Next, we consider the virtual work due to the exerted control torques on the joints, which is given by $A_i^\top\tau_i$ in the wing frame. There will be a reactive torque, namely $(-\tau_i)$ exerted to the body. The total corresponding virtual work will be
\begin{align}
    \delta \mathcal{W}_{\tau}=\sum_{i=1}^4\left[(A_i^\top \tau_i)^\top \eta_i + (-\tau_i)^\top \eta\right].
\end{align}

Now, we replace the aerodynamic forces and torques given by \eqref{eq10}-\eqref{eq12} in \eqref{eqwi}, in view of the fact that $\delta \mathcal{W}=\delta \mathcal{W}_{\tau}+\sum_{i=1}^4\delta \mathcal{W}_i$, the right-hand side of \eqref{alembert} yields
\begin{align*}\label{rhs}
\begin{split}
    &\int_{t_{0}}^{t_{f}}\delta \mathcal{W}dt=\int_{t_{0}}^{t_{f}}\Bigg\{  \left(A_B\sum_{i=1}^{4} A_{i} F_{i}\right)^\top\delta p\\
    &+\left(\sum_{i=1}^{4} \left(\mu_{i} \times A_{i} F_{i}-\tau_{i}\right)\right)^\top\eta
    + \sum_{i=1}^{4}\left(M_{i}+A_{i}^\top \tau_{i}\right)^\top\eta_{i}\Bigg\} dt.
    \end{split}
\end{align*}
 The Lagrange-D'Alembert equations are obtained by grouping the terms from the right-hand side, and the left-hand side that are multiplied by the linearly independent infinitesimal variations $\delta x$, $\delta \eta$ and $\delta \eta_i$. Then,  invoking Hamilton's principle that states that $\delta \mathfrak{S}=0$, for all possible variations with fixed endpoints, and using the fact that the infinitesimal variations vanish at $t_0$  and $t_f$, one obtains the Lagrange-D'Alembert equations given in \eqref{eq17}-\eqref{eq19} as a result of the value inside of the integral in equations \eqref{lhs}, and \eqref{rhs} being equal to zero

\section{The reduced Dynamical Model}\label{apen4}
To obtain the reduced dynamical model presented in \eqref{eq31}, first, we will proceed by writing the matrices $\mathbf{C}, \textbf{N}$ and $S$, and the vectors  $F_u:=H_c\tau, F_{\text{g}}$ and $F_{a}$,  presented in equation \eqref{eq28} as follows:
\begin{fleqn}

\begin{align*}
    \mathbf{C}=\begin{bmatrix}
        \tilde{\mathbf{C}}_{11} &\tilde{\mathbf{C}}_{12}\\\tilde{\mathbf{C}}_{21}&\tilde{\mathbf{C}}_{22}   
    \end{bmatrix}, \quad \mathbf{N}=\begin{bmatrix}
        \tilde{\mathbf{N}}_{11} &\tilde{\mathbf{N}}_{12}\\\tilde{\mathbf{N}}_{21}&\tilde{\mathbf{N}}_{22},   
    \end{bmatrix},\quad S=\begin{bmatrix}
        \tilde{S}_{11}&\tilde{S}_{12}\\
        \tilde{S}_{21}&\tilde{S}_{22}
    \end{bmatrix}
\end{align*}

\begin{align*}
   F_a=\begin{bmatrix}
        \tilde{F}_{a1}\\\tilde{F}_{a2}
    \end{bmatrix},\quad F_u=\begin{bmatrix}
        \tilde{F}_{u1}\\\tilde{F}_{u1}
    \end{bmatrix}, \quad F_{\text{g}}=\begin{bmatrix}
        \tilde{F}_{\text{g}1}\\\tilde{F}_{\text{g}1}
    \end{bmatrix},
\end{align*}
\end{fleqn}
where the matrices $\tilde{\mathbf{C}}_{11}, \tilde{\mathbf{N}}_{11} $, $\tilde{S}_{11} \in \mathbb{R}^{6 \times 6}$, and $\tilde{\mathbf{C}}_{12}, \tilde{\mathbf{N}}_{12} $, $\tilde{S}_{12} \in \mathbb{R}^{6 \times 12}$, and $\tilde{\mathbf{C}}_{21}, \tilde{\mathbf{N}}_{21} $, $\tilde{S}_{21} \in \mathbb{R}^{12 \times 6}$,  and $\tilde{\mathbf{C}}_{22}, \tilde{\mathbf{N}}_{22} $, $\tilde{S}_{22} \in \mathbb{R}^{12 \times 12}$. The vectors $\tilde{F}_{a1}, \tilde{F}_{u1},$ $\tilde{F}_{\text{g}1} \in \mathbb{R}^6$, and  $\tilde{F}_{a2}, \tilde{F}_{u2}$, $\tilde{F}_{\text{g}2} \in \mathbb{R}^{12}$. Moreover, from the expression of $F_u:=H_c\tau$, one can easily notice the following relationship
\begin{align}
    \tilde{F}_{u1}= \begin{bmatrix}
    0 &0 &0&0\\ -A_1&-A_2&-A_3&A_4
    \end{bmatrix}\tilde{F}_{u2}=:K\tilde{F}_{u2}.
\end{align}
Replacing $\tilde{F}_{u2}$ with the wings dynamics given by the $3$th, $\dots,$ $6$th rows of  \eqref{eq28}, one can write the translational and rotational dynamics of the main body as follows:

   \begin{align}
    \begin{split}
        &\left(\tilde{\mathbf{C}}_{11}-K\tilde{\mathbf{C}}_{21}\right)\dot{\xi}_B=\left(K\tilde{S}_{22} \tilde{\mathbf{C}}_{21}-\tilde{S}_{11}\tilde{C}_{11}\right)\xi_B\\
        &+\left(K\tilde{\textbf{N}}_{21}-\tilde{\textbf{N}}_{11}\right)\xi_B+\left(K\tilde{\mathbf{C}}_{22}-\tilde{\mathbf{C}}_{12}\right)\dot{\xi}_w\\
        &+\left(K\tilde{S}_{22}\tilde{\mathbf{C}}_{22}-\tilde{S}_{11}\tilde{\mathbf{C}}_{12}\right)\xi_w+\left(K\tilde{\textbf{N}}_{22}-\tilde{\textbf{N}}_{12}\right)\xi_w\\
        &+\tilde{F}_{a1}+\tilde{F}_{g1}-K\left(\tilde{F}_{a2}+\tilde{F}_{g2}\right).
        \end{split}
    \end{align}
Consequently, the matrices
$\mathcal{C}_t^{\Theta}$ and $\mathcal{D}_t^{\Theta}$ and the vectors $\mathcal{V}_t^{\Theta}, \mathcal{F}_t^{\Theta}$ in \eqref{eq31} are given by  
\begin{align*}
    &\mathcal{C}^t_\Theta(\mathfrak{g}_B)= \left(\tilde{\mathbf{C}}_{11}-K\tilde{\mathbf{C}}_{21}\right),\\
    &\mathcal{D}^t_\Theta(\mathfrak{g}_B,\xi_B)=\Bigg(\left(K\tilde{S}_{22} \tilde{\mathbf{C}}_{21}-\tilde{S}_{11}\tilde{\mathbf{C}}_{11}+K\tilde{\textbf{N}}_{21}-\tilde{\textbf{N}}_{11}\right)\\
    &+\begin{bmatrix}
        0\\ \displaystyle\sum_{j \in\{1,\dots,4\}} A_j\displaystyle\sum_{i=1}^4\left(\widehat{J_i\Omega_i}A_i^\top-m_i\widehat{\hat{\kappa}_i\Omega_i}A_i^\top\hat{\mu}_i\right)
    \end{bmatrix}\\
    &+\begin{bmatrix}
        0\\
        \displaystyle\sum_{i=1}^4m_iA_B\widehat{A_i\hat{\kappa}_i\Omega_i}
    \end{bmatrix}+\begin{bmatrix}
        0\\\displaystyle\sum_{i=1}^4-\widehat{J_i^{[23]}\Omega_i}
    \end{bmatrix}\Bigg)\\
&\mathcal{V}^t_\Theta\left(\mathfrak{g}_B\right)=\left(K\tilde{\mathbf{C}}_{22}-\tilde{\mathbf{C}}_{12}\right)\dot{\xi}_w+\Big(K\tilde{S}_{22}\tilde{\mathbf{C}}_{22}-\tilde{N}_w\Big)\xi_w+\\
& \tilde{F}_{\text{g}1}-K\tilde{F}_{\text{g}2}\\ 
&\mathcal{F}^t_\Theta\left(\mathfrak{g}_B,\xi_B\right)=\tilde{F}_{a1}-K\tilde{F}_{a2}.
\end{align*}
with $\tilde{N}_w$ defined as follows
\begin{multline*}
    \tilde{N}_w:=\\\begin{bmatrix}
m_1A_BA_1\hat{\Omega}_1\hat{\kappa}_1&m_iA_BA_i\hat{\Omega}_i\hat{\kappa}_i&m_iA_BA_i\hat{\Omega}_i\hat{\kappa}_i&m_iA_BA_i\hat{\Omega}_i\hat{\kappa}_i\\\mathbf{N}_{23}&\mathbf{N}_{24}&\mathbf{N}_{25}&\mathbf{N}_{26}
     \end{bmatrix}.
\end{multline*}

\begin{table}
\centering
    \begin{tabular}{c|c}
        Parameter  & Numerical Value  \\
       \hline $\rho$ & $1.2 \mathrm{kg/m^3}$\\
$m_B$ &  $5.4483e^{-5}\mathrm{kg} $ \\
$m_{1,2}$ & $1.9069e{-6}\mathrm{kg} $ \\
$m_{3,4}$ & $2.3126e^{-6}\mathrm{kg}$ \\
$l_{1,2}$ & $0.0185\mathrm{m} $  \\
$l_{3,4}$ & $0.025 \mathrm{m}$\\
\hline
    \end{tabular}
    \caption{Wings morphology parameters.}
    \label{tab:morphology}
\end{table}


\section{Dragonfly Morphology}\label{apen6}
The dragonfly morphological parameters are calculated from the dead dragonfly shown in Fig. ~\ref{fig22}. First, we will suppose that the main body is a cylinder, with a diameter of $D_B=7.6$mm and a height of $H_B=25$mm. Using this assumption, and the fact that the distance between the hind and fore wings' roots is $5.42$mm, one can easily calculate $J_B$ and $\mu_i$ as follows:
\begin{align}
    &J_B=10^{-8}\begin{bmatrix}
    0.0109         &0        & 0\\
         0    &0.2847  &       0\\
         0       &  0   & 0.2847\\
\end{bmatrix},
\end{align}
\begin{align}
    \begin{split}
        &\mu_1=10^{-3}\begin{bmatrix}
            2.71& 3.8& 0
        \end{bmatrix}^\top, \quad \mu_2=10^{-3}\begin{bmatrix}
            2.71 &-3.8& 0
        \end{bmatrix}^\top,\\
        &\mu_3=10^{-3}\begin{bmatrix}
            -2.71& 3.8 &0
        \end{bmatrix}^\top, \quad \mu_4=10^{-3}\begin{bmatrix}
            -2.71& -3.8 &0
        \end{bmatrix}^\top.\\
    \end{split}
\end{align}
Next, using the wings' polynomials $q_i^{LE}$ and $q_i^{TE}$, one can directly calculate the wings parameters $J_i, \kappa_i, m_i$ and $l_i$ as follows: 
\begin{align}
\begin{split}
        &\kappa_{1,2}=10^{-3}\begin{bmatrix}
            7.978& (-1)^{i+1}4.975&0
        \end{bmatrix}^\top, \\
        &\kappa_{3,4}=10^{-3}\begin{bmatrix}
            1.175 &(-1)^{i+1}5.89& 0
        \end{bmatrix}^\top,
        \end{split}
\end{align}
\begin{align*}
    &J_{1,2}=10^{-3}\begin{bmatrix}
        0.2303 &  (-1)^{i+1}0.1902&   0\\
   (-1)^{i+1}0.1902   & 0.1731       &  0\\
         0      &   0 &   0.4034
    \end{bmatrix},
\end{align*}

\begin{align*}
    J_{3,4}=10^{-3}\begin{bmatrix}
        0.4164   & (-1)^{i+1}0.0693&         0\\
   (-1)^{i+1} 0.0693   & 0.0326       &  0\\
         0    &     0  &  0.4489
    \end{bmatrix}.
\end{align*}


\section{The Control Dynamical Model}\label{apen5}

The control dynamical model, presented in \eqref{eq32}, is obtained by considering the aerodynamic forces and torques as control inputs denoted by $F_C$ and $\Gamma_C$. The terms that represent the inertial forces and torques due to the body motion are grouped and denoted by $F_B$ and $\Gamma_B$ \textit{i.e.} the terms that are multiplied by $\dot{\xi}_B$ and $\xi_B$. Finally, the terms that are multiplied by $\dot{\xi}_w$ and $\xi_w$ alone, represent the forces and torques due to the wing motion.
Consequently, the forces and torques expressions are given by


   
\begin{align}
 F_c&=\displaystyle\sum_{i=1}^4A_BA_iF_i,\\
\nonumber F_w&=\displaystyle\sum_{i=1}^4\Big(2m_iA_B\hat{\Omega}_BA_i\hat{\kappa}_i\Omega_i+m_iA_i\hat{\kappa}_i\dot{\Omega}_i+\\
      &\qquad m_iA_i\hat{\Omega}_i\hat{\kappa}_i\Omega_i\Big) \\
    \nonumber F_B&=\displaystyle\sum_{i=1}^4\Bigg(m_iA_B\left(\hat{\mu}_i+\widehat{A_i\kappa_i}\right)\dot{\Omega}_B+\\
    &\qquad m_iA_B\hat{\Omega}_B\left(\hat{\mu}_i+\widehat{A_i\kappa_i}\right)\Omega_B\Bigg),\\
    \Gamma_c&=\displaystyle\sum_{i=1}^4\left(\hat{\mu}_iA_iF_i+A_iM_i \right).
\end{align}

\begin{fleqn}

\begin{strip}
\hrulefill
\begin{equation}
\begin{split}
\Gamma_w=&-\sum_{i=1}^4\Bigg(A_i\hat{\Omega}_iJ_iA_i^\top+A_iJ_i\hat{\Omega}_i^\top A_i^\top+m_i\left(\hat{\mu}_i^\top\widehat{A_i\hat{\Omega}_i\kappa_i}+\widehat{A_i\hat{\Omega}_i\kappa_i}^\top \hat{\mu}_i\right)-m_iA_i\hat{\kappa}_i\hat{\Omega}_iA_i^\top\hat{\mu}_i\Bigg)\Omega_B\\&+ \sum_{i=1}^4\Bigg(A_i\widehat{A_i^\top\Omega_B}J_i+m_i\widehat{A_i^\top\hat{\mu}_i\Omega_B}\hat{\kappa}_i+m_i\hat{\Omega}_B\hat{\mu}_iA_i\hat{\kappa}_i+
\hat{\Omega}_BA_iJ_i\Bigg)\Omega_i+\sum_{i=1}^4\bigg(m_iA_i\hat{\Omega}_i\hat{\kappa}_i^\top A_i^\top\hat{\mu}_i\Omega_B\\
&-\left(2A_iJ_i-m_i\hat{\mu}_iA_i\hat{\kappa}_i\right)\dot{\Omega}_i\bigg).
\end{split}
\end{equation}
    
\begin{equation}
\begin{split}
  \Gamma_B=&-\sum_{i=1}^4m_i\left(\hat{\mu}_i+2\widehat{A_i\kappa_i}\right) A_B^\top\Ddot{p}
    -\sum_{i=1}^4 \Bigg(m_{i} \hat{\mu}_{i}^\top \hat{\mu}_{i}+2A_{i} J_{i} A_{i}^\top+m_{i}\left(\hat{\mu}_{i}^\top \widehat{A_{i} \kappa_{i}}+2\widehat{A_{i} \kappa_{i}}^\top \hat{\mu}_{i}\right)\Bigg)\dot{\Omega}_B\\
    &-\sum_{i=1}^4\Bigg(-m_i\left(\hat{\mu_i}+\widehat{A_i\kappa_i}\right)\hat{\Omega}_BA_B^\top+m_i\left(\widehat{\hat{\mu}_i\Omega_B}+\widehat{\widehat{A_i\kappa_i}\Omega_B}\right)A_B^\top+m_i\hat{\Omega}_B\left(\hat{\mu}_i+\hat{A_i\kappa_i}\right)A_B^\top\Bigg)\dot{p}\\
&-\sum_{i=1}^4\Bigg(-A_i\widehat{J_iA_i^\top\Omega_B}A_i^\top-m_iA_i\hat{\kappa}_iA_i^\top\left(\hat{\Omega}_B\hat{\mu}_i-A_i\widehat{\hat{\mu}_i\Omega_B}\right)+m_i\hat{\Omega}_B\left(\hat{\mu}_i+\widehat{A_i\kappa_i}\right)A_B^\top\Bigg) \Omega_B+\sum_{i=1}^4m_i\text{g}\hat{\mu}_iA_B^\top e_3.
\end{split}
\end{equation}
\hrulefill
\end{strip}
\end{fleqn}



\end{document}